\documentclass[useAMS, usenatbib]{mn2e}
\usepackage{times}
\usepackage{amsmath}
\usepackage{xspace}
\usepackage{psfig}
\usepackage{defs}

\def\eq{equation}
\def\fig{Fig.}

\def\tab{table}

\def\cf{{\it cf.}}
\def\ie{{\it i.e.}}
\def\eg{{\it e.g.}}
\def\ltsima{$\; \buildrel < \over \sim \;$}
\def\simlt{\lower.5ex\hbox{\ltsima}}
\def\gtsima{$\; \buildrel > \over \sim \;$}
\def\simgt{\lower.5ex\hbox{\gtsima}}
\def\fesc{{$\langle f_{\rm esc}\rangle$}\xspace}
\def\h2{H$_2$\xspace}
\def\H2{H$_2$\xspace}
\def\m{$^{-1}$\xspace}

\def\mmm{$^{-3}$\xspace}
\def\pp{$^2$\xspace}

\def\ion#1#2{\text{#1\,\sc #2}}
\def\HI{{\ion{H}{i} }}
\def\HII{{\ion{H}{ii} }}
\def\GI{{\ion{He}{i} }}
\def\GII{{\ion{He}{ii} }}
\def\GIII{{\ion{He}{iii} }}

\def\popI{Population~I\xspace}
\def\popII{Population~II\xspace}
\def\pop3{Population~III\xspace}

\def\p3{``small-halo''\xspace}
\def\pp3{``Small-halo''\xspace}

\def\lya{Lyman-$\alpha$\xspace}

\def\taue{$\tau_{\rm e}$\xspace}
\def\sig8{$\sigma_8$\xspace}
\def\zrei{$z_{\rm rei}$\xspace}
\def\xe{$\langle x_{\rm e} \rangle$\xspace}
\def\bi{\begin{itemize}}
\def\ei{\end{itemize}}

\title[X-ray Preionisation. I]{X-ray Preionisation Powered by Accretion
  on the First Black Holes. I: a Model for the WMAP Polarisation Measurement}

\author[M. Ricotti and J.P. Ostriker]
{Massimo Ricotti and Jeremiah P. Ostriker\\ 
Institute of Astronomy, Madingley Road, Cambridge CB3 0HA\\
ricotti@ast.cam.ac.uk, jpo@ast.cam.ac.uk}
\date{Accepted ---. Received ---; in original form 31 October 2003}
\pagerange{\pageref{firstpage}--\pageref{lastpage}}
\pubyear{2002}
\begin{document}
\maketitle
\label{firstpage}

\begin{abstract}
  In this paper we investigate the possibility that there is a first
  phase of partial ionisation due to X-rays produced by black hole
  accretion in small-mass galaxies at redshifts $7 < z < 20$. This is
  followed by complete reionisation by stellar sources at $z \simeq
  7$. This scenario is motivated by the large optical depth to
  Thompson scattering, \taue$\simeq 0.17 \pm 0.04$, recently measured
  by WMAP. But it is also consistent with the observed Gunn-Peterson
  trough in the spectra of quasars at $z \sim 5-6$. We use a
  semianalytic code to explore models with different black hole
  accretion histories and cosmological parameters.  We find that
  ``preionisation'' by X-rays can increase the intergalactic medium
  (IGM) optical depth from \taue$\approx 0.06$ given by stellar
  sources only, to $0.1 \simlt \tau_{\rm e} \simlt 0.2$, if a fraction
  of baryons $10^{-5} \simlt \omega_{\rm ac} \simlt 10^{-4}$ is
  accreted onto seed black holes produced in the collapse of low
  metallicity, high mass stars before $z \simeq 15$.  To be effective,
  preionisation requires a non-negligible star formation in the first
  small-mass galaxies in which seed black holes are formed.  By $z
  \sim 20-25$ the IGM is re-heated to $10,000$ K and the ionisation
  fraction is about 20\%.  The increase of the IGM Jeans mass is
  effective in reducing star formation in the smaller-mass haloes.
  Large values of \taue are obtained in models with top-heavy stellar
  initial mass function only if pair-instability supernovae are not
  important. Seed black holes are assumed to accrete at near the
  Eddington limit with a duty cycle that decreases slowly with
  increasing time.  Alternatively, a moderate fraction of the black
  holes must be ejected from the host galaxy or exist without merging
  into the supermassive black holes in galactic centres.  The model
  predicts that dwarf spheroidal galaxies, if they are preserved
  fossils of the first galaxies, may host a mass in black holes that
  is 5-40\% of their stellar mass. The redshifted X-ray background
  produced by this early epoch of black hole accretion constitutes
  about $5-10$\% of the X-ray background in the 2-50 keV bands
  and roughtly half of the currently estimated black hole mass density
  was formed at early times.  Moreover, in most models, the photons
  from the redshifted background are sufficient to fully reionise \GII
  at redshift $z \sim 3$ without any additional contribution from
  quasars at lower redshifts and the temperature of the mean density
  intergalactic medium remains close to $10^4$ K down to redshift $z
  \sim 1$.
\end{abstract}

\section{Introduction}

The mean transmitted flux along the line of sight of the furthest
quasars found by the Sloan Digital Sky Survey
\citep[\eg,][]{Becker:01} shows that the ionisation fraction of the
intergalactic medium (IGM) is rapidly decreasing approaching $z \sim
6.3$. Assuming, at that redshift, an approximately constant stellar
emissivity of ionising radiation, the IGM would have been reionised at
\zrei$\sim 7$ by stellar sources \citep[][ and
others]{Gnedin:00,Djorgovski:01,Songaila:02,Fan:03}.  But a much
earlier ionisation epoch is implied by the WMAP observations
\citep{Bennet:03, Kogut:03}. If the IGM had a sudden transition from
neutral to completely ionised, producing the optical depth to Thompson
scattering \taue$=0.17 \pm 0.04$ measured by WMAP, \zrei$ \simeq 17$
for the best fit $\Lambda$CDM cosmological model \citep{Spergel:03}.

In a companion paper \citep[paper~I,][]{RicottiOI:03}, we have tried to
explain these apparently contradictory results by considering the
contribution to reionisation due to zero-metallicity (\pop3) stars.
Since such stars are thought to be massive ($M_* > 5-10$ M$_\odot$)
\citep[\eg,][]{OmukaiN:98, Abel:00, Bromm:02, Abel:02, NakamuraU:02} and
hot \citep{Tumlinson:00}, their efficiency for UV emission is larger
than for \popII and \popI stars that have a Salpeter initial mass
function (IMF).  A time dependent efficiency of UV emission,
$\epsilon_{\rm UV}$, produced by the transition from a top-heavy IMF
at high redshift to a Salpeter IMF at low redshift, could possibly
explain the large \taue measured by WMAP and would be consistent with
the Sloan data at $z \simlt 6.3$ \citep{ChiuFO:03}.  An upper limit of
the UV emissivity, $\epsilon_{\rm UV}^{\rm max} \simlt 2-3\times
10^{-3}$ can be derived by assuming the maximum efficiency of energy
production by thermonuclear reactions in massive stars. For a Salpeter
IMF, $\epsilon_{\rm UV} \sim 2-4 \times 10^{-4}$, depending on the
stellar metallicity \citep{Tumlinson:00}.  Therefore, assuming a
constant (maximal) escape fraction of ionising photons from the
galaxies, \fesc$\simeq 1$, a redshift dependent IMF would increase the
UV emission efficiency from low to high redshift by about 10-20 times.

Several authors have tried to reproduce the high optical depth
measured by WMAP with the increased ionising photon emission by
zero-metallicity stars. \cite{Cen:03b} has used a semianalytic
calculation to indicate that the IGM may have experienced two distinct
epochs of reionisation. \cite{HaimanH:03} have investigated different
reionisation histories showing that optical depths consistent with
WMAP can be achieved assuming very massive \pop3 stars.
\cite{WyitheL:03} show that, if feedback regulates star formation in
early low-mass galaxies as observed in present-day dwarfs, \pop3 stars
forming with a heavy IMF are required to match WMAP data.
\cite{CiardiFW:03}, using numerical simulation, show that zero
metallicity stars with a moderately heavy IMF (characteristic stellar
mass of 5 M$_\odot$) and \fesc$=20$\% are able to generate an optical
depth of 0.17, consistent with WMAP data. \cite{Sokasian:03}, also
using numerical simulations, showed that zero metallicity stars with a
top-heavy IMF and \fesc$\sim 100$\% is needed to produce an optical
depth consistent with WMAP data.

In paper~I we have found (consistent with the work of the other
investigators mentioned above) that, assuming the most extreme
properties for zero metallicity stars (top heavy IMF and \fesc$\sim
100$\%), the maximum optical depth to Thomson scattering produced by
\pop3 stars is \taue$ \sim 0.13$, marginally consistent with WMAP
data. In order to produce this large value of \taue, massive zero-metallicity 
(\pop3) stars must be the dominant population until redshift $z \simeq 10$.
But we have noted that, if the ratio of metal atoms to ionising
photons produced by \pop3 stars is normal, the metal enrichment of the
ISM prevents metal-poor stars from being produced for long enough to
ionise more than a small fraction of the IGM, \ie that the \pop3 phase
is very rapidly self-limiting. In addition, if pair-instability SNe
are important, the mechanical energy input by SN explosions produces
strong outflows in galaxies with masses $M_{\rm dm} \simlt 10^{9}$
M$_\odot$, reducing their star formation and delaying reionisation.
These two effects will lower the redshift of reionisation and \taue to
values inconsistent with WMAP but still consistent with the Sloan
quasar data. Thus we have found it essentially impossible to reproduce
the WMAP and Sloan results using stellar sources of UV ionisation.

In summary, the reason for the partial disagreement with the
conclusions of previous works on the ability of \pop3 stars to produce
the large optical depth observed by WMAP, is motivated by our findings
concerning the self-termination of \pop3 stars by the enhanced metal
pollution and the negative feedback on star formation from the
enhanced energy input by SN explosions, that inevitably follow from
the assumption of a top-heavy IMF. In previous works these concerns
have not been addressed.  However in paper~I we also note that, if
most \pop3 stars collapse into black holes (BHs), the metal pollution
problem is alleviated and the mechanical feedback from SN/hypernova
explosions will be reduced as well.  An interesting consequence of
this scenario is a copious production of BHs from the first stars
\citep{Schneider:02}.  For this reason it is possible that the
secondary radiation from accretion onto seed BHs might be a more
important source of ionising radiation than the primary radiation from
the same stars. In particular because, as we will show, accretion on
seed BHs is not sensitive to the early termination of the \pop3 epoch
by metal pollution.

Motivated by these results, in this paper we study the partial
ionisation of the IGM by an early X-ray background produced by
accreting BHs, followed by more complete stellar reionisation by
\popII stars at \zrei$ \approx 7$. An hybrid model where UV from \pop3
stars and X-ray emission from accreting BHs are both important in
producing a large optical depth is also viable if self-termination of
\pop3 stars happens late (\eg, $z \sim 10$) and if we assume extreme
parameters for \fesc and for the global \pop3 stars UV emissivity.
But, given the difficulties in modelling properly the transition from
\pop3 to \popII stars, the predicted importance of UV from \pop3 stars
is very uncertain (almost a free parameter at this point). As an
example, in the present paper we show a model where the transition
from \pop3 to \popII stars is at redshift $z \sim 15$. In this case we
show that UV radiation from \pop3 stars has negligible effect on the
Thompson optical depth.

For a fixed emissivity of ionising radiation, X-rays are less
efficient in reionising the IGM, because most of their energy goes
into heat instead of ionisation. The inclusion of secondary electrons
can boost the ionisation efficiency by about a factor of ten if the
electron fraction of the IGM is \xe$<10$ \%, but roughly $1/3$ of the
primary electron energy is always converted into heating of the IGM
\citep{ShullVan:85}. It will be shown that the partial ionisation of
the IGM must begin at redshifts $z \simgt 15$ in order to have an
important effect on \taue.  This can only happen if the X-ray sources
(\ie, accreting BHs) form in small-mass haloes with masses $M_{\rm dm}
\simlt 10^8$ M$_\odot$.  Previous work on the formation of the first
galaxies \citep{RicottiGSa:02, RicottiGSb:02} has used cosmological
simulations with radiative transfer to show that star formation in the
first small-mass galaxies is reduced by feedback effects but is not
suppressed. It is therefore plausible that a substantial production of
seed BHs from the first stars took place in the first small-mass
galaxies.

Published studies have investigated the effects of X-rays on early
galaxy formation using a semianalytic approach \citep{Oh:00,
  Venkatesan:01} and cosmological simulations \citep{Machacek:03}.
But since those works were completed before WMAP results, they did not
focus on scenarios that could produce \taue$\simeq 0.17$.  The study
presented in this paper (paper~IIa) is carried out using a semianalytic
code based on principles similar to those adopted by \cite{Chiu:00}.
In \cite*{RicottiOG:03} (paper~IIb) we present the results of
cosmological hydrodynamic simulations that include radiative
transfer. We discuss the observational signatures of X-ray
preionisation compared to stellar reionisation, including calculations
of the expected amplitude of secondary anisotropies of the CMB and the
redshifted 21cm signal in absorption and emission against the CMB. The
semianalytic models presented in this paper allow us to explore a
larger parameter space and help us to interpret the numerical result
of the full cosmological simulations. This new code, presented in
Appendix~\ref{ap:A}, calculates the evolution of the filling factor of
\HII regions and the ionisation and thermal history of the IGM outside
the \HII regions, produced by an X-ray background.  The semianalytic
code has been tested and calibrated using the results of the
cosmological simulations (\cf, paper~I). Note that some cosmological
simulations presented here include the effects of SN explosions, using
the recipe discussed in \cite{Gnedin:98a}, in contrast to earlier
papers in this series \citep{RicottiGSa:02, RicottiGSb:02}.

This paper is organised as follows. In \S~\ref{sec:rat} we present
qualitative arguments showing why and for which models X-ray
preionisation can be more effective in increasing \taue than
reionisation by stellar sources. In \S~\ref{sec:xsour} we estimate the
accretion rate onto seed BHs and we compute the integrated X-ray
energy required to explain the large \taue measured by WMAP. In
\S~\ref{sec:res_sa} we summarise the characteristics of the
semianalytic code for reionisation and we show its results.
In \S~\ref{sec:conc} we conclude this work with a summary and a
discussion of some of the observable effects of early black hole X-ray
heating.
A full description of the semianalytic code is given in
Appendix~\ref{ap:A}.

\section{X-Ray preionisation: rationale}\label{sec:rat}

\begin{figure}
\centerline{
\psfig{figure=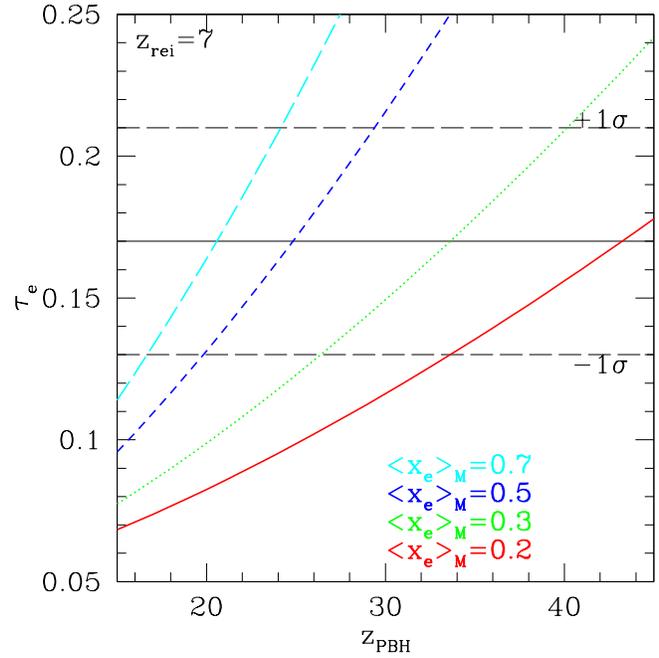,width=9cm}
}
\caption{\label{fig:ske1} Optical depth to Thomson scattering, \taue,
  for a toy model where a complete reionisation at redshift \zrei$=7$
  is preceded by constant, partial ionisation \xe, starting at
  redshift $z=z_{\rm PBH}$.  The lines from bottom to the top refer to
  mean (mass weighted) ionisation fractions \xe$_M=0.2, 0.3, 0.5$, and
  $0.7$.
}
\end{figure}

The essential physics is easily understood. Ultraviolet (UV) photons
from stars at wavelengths only moderately shorter than $912$~\AA~
($h\nu=13.6$ eV) have a very short mean free path in neutral hydrogen.
As a consequence they produce Str\"omgren spheres surrounding the UV
sources in high density regions. In such regions recombination rates
are high and many photons are required to keep one hydrogen atom
ionised for a Hubble time. The much larger volumes between the
Str\"omgren spheres remain sensibly neutral. However, much harder
photons (\eg, 1 keV) can uniformly fill space and thus lead to
fractional ionisation which is higher in the low-density regions than
in the high-density regions containing the sources. But, as noted
earlier, the fractional ionisation is low but the volume is large
since the volume fraction in virialised haloes at $z=15$ is only a few
per cent.  To illustrate this let us show how a simple {\it ad hoc}
ionisation history affects \taue.  In \fig~\ref{fig:ske1}, we show the
IGM optical depth to Thompson scattering, \taue, in a model where a
complete reionisation at redshift \zrei is preceded by a partial
ionisation, with constant ionisation fraction \xe, starting at
$z=z_{\rm PBH}$. The long-dashed, dashed, dotted and solid lines in
\fig~\ref{fig:ske1} show \taue as a function of $z_{\rm PBH}$ assuming
complete reionisation at \zrei$=7$ and preionisation electron fraction
\xe$=0.7, 0.5, 0.3$ and 0.2, respectively. We see that two conditions
need to be met in order to have \taue$\simeq 0.17$: (i) preionisation
must start early, at $z_{\rm PBH} \sim 20-40$, and (ii) the IGM
preionisation fraction must be \xe$\simgt 20$\%.

Condition (i) is met only if the X-ray sources can form in haloes with
masses $M_{\rm dm} \simlt 10^8$ M$_\odot$. The two peculiarities of
these small-mass haloes are that they rely on molecular hydrogen
cooling to form stars, and their dark matter gravitational potential
is too shallow to retain gas hotter than $T \approx 2 \times 10^4$ K.
For these reasons small-mass galaxies are subject to radiative
feedback effects that might suppress their formation (\eg,
\citealp{HaimanRL:97, Machacek:03}, but see \citealp{RicottiGS:01},
who find a new positive feedback effect).  Using cosmological
simulations with radiative transfer \cite{RicottiGSa:02,
  RicottiGSb:02} found that the main negative feedback thought to
suppress the formation of the first galaxies (the \H2
photodissociating background) is not the dominant effect.  Instead,
feedback by ionising radiation plays the dominant r\^ole, inducing a
bursting star formation mode in small-mass galaxies.  The feedback
mechanism is complex because it is composed of several processes, each
with different relevance depending on the parameters of the simulation
(mainly the ionising escape fraction, \fesc, and the IMF). In brief,
photoevaporation and \H2 formation/photodissociation produce a
bursting mode of star formation that is catalyst for molecular
hydrogen re-formation inside relic \HII regions and in the precursors
of cosmological Str\"omgren spheres.  Star formation in small-mass
galaxies is self-regulated on a cosmological scale and it is reduced
by radiative feedback but is not suppressed. Even if the volume
fraction filled with ionised Str\"omgren spheres never exceeds a few
per cent, these small-mass galaxies form enough \pop3 stars and
therefore seed black holes (see \S~\ref{sec:xsour}) to produce, via
accretion, a substantial X-ray background at high redshift.

Condition (ii) can be met if the X-ray energy input is sufficiently
large. Note that ionisation by secondary electrons becomes
increasingly inefficient when \xe$ \simgt 10-20$\% since the secondary
electrons loose an increasing fraction of their energy to heat and a
decreasing fraction to further ionisation as the neutral fraction
declines. We can estimate the fraction of baryons, $\omega_{\rm ac}$,
that we need to accrete by counting the number of ionisations per
hydrogen atom, $N_{\rm Xph} \simeq$\xe, needed to partially ionise the
IGM to a level \xe$\sim 50$\% (the fractional ionisation of the IGM
must be \xe$\simgt 20$\% in order to produce the large optical depth
to Thomson scattering measured by WMAP, \cf~\fig~\ref{fig:ske1}). If
we neglect recombinations we have
\begin{equation}
N_{\rm Xph} = 7.3 \times 10^7 \left(\epsilon_{\rm X} \over
  X_{\rm Xsec}
  \right)\omega_{\rm ac} \simeq \langle x_{\rm e} \rangle \simeq 0.5,
\label{eq:nphx}
\end{equation}
where $\epsilon_{\rm X}$ is the energy emitted in the X-ray band per
hydrogen atom accreted on the BHs and $X_{\rm Xsec}$ is a pure number
defined as $X_{\rm Xsec}=X_{\rm X}/f_{\rm sec}\sim 20$, where $f_{\rm
  sec} \sim 5-10$ takes into account the extra ionisation by secondary
electrons (we have estimated this number switching off secondary
ionisations in our semianalytic code, presented in the appendix), and
$X_{\rm X}=\overline{h \nu} /(13.6 {\rm eV})$ is the mean energy of
X-ray photons in units of the \HI ionisation energy for the assumed X-ray
spectrum.  For accretion around a BH in which the gas is highly ionised and
electron scattering is the most important opacity, the isotropic
luminosity produced by the accretion cannot exceed the Eddington
luminosity
\begin{equation}
L_{\rm Ed} \equiv {4\pi GM_{\rm BH} m_p \over \sigma_T}=(1.3 \times 10^{38} {\rm erg~s}^{-1}) \left({M_{\rm BH} \over M_\odot}\right),
\end{equation}
where $\sigma_T$ is the Thomson scattering cross section.  Given the
accretion efficiency $\epsilon =L/\dot M_{\rm ac}c^2\sim 0.15$,
typical for quasars, the mass accreted at the Eddington rate is $\dot
M_{\rm Ed}=L_{\rm Ed}/\epsilon c^2$ and the timescale of the accretion is
the Eddington time $t_{\rm Ed} =M_{\rm BH}/\dot M_{\rm Ed} \sim 10^8$
yrs.

Observations of active galactic nuclei (AGN) at low-redshift show
that, when the central BH is accreting, it does so at near the
Eddington limit and, when it is quiet, the energy output is
negligible.  In \S~\ref{sec:xsour} we will introduce a parameter,
$f_{\rm duty}$, to model the observed duty cycle of AGN. We define
$f_{\rm duty}$ as the fraction of time when the BH is accreting at the
Eddington limit.  Assuming a spectral energy distribution (SED)
typical for AGNs (\cf, \fig~\ref{fig:sed}), a fraction $\beta=L_{\rm
  X}/L_{\rm Ed} \sim 10$\% of the energy output from the accreting BH
is emitted in the X-ray band, with a resulting efficiency of X-ray
production $\epsilon_{\rm X}=\beta \epsilon$.  Finally, using
\eq~(\ref{eq:nphx}), we can estimate the fraction of baryons that
needs to be accreted to partially reionise the IGM,
\begin{equation}
\omega_{\rm ac} \simeq 6.8 \times 10^{-6} \left({0.02 \over
    \epsilon_{\rm X}}\right) \left({X_{\rm Xsec} \over
    20}\right).
\label{eq:omBH}
\end{equation}
As a comparison we also estimate the baryon fraction needed to be
converted into stars to reionise the IGM.  The maximum radiative
efficiency from thermonuclear processes is $\epsilon_{\rm UV}^{\rm
  max} =2-3 \times 10^{-3}$.  Since UV photons have a short mean free
path they first ionise, and keep ionised, the high density regions
around each source before escaping into the IGM. In this case, then we
cannot neglect recombinations. The number of UV photons per baryon
needed to reionise the IGM is $N_{\rm UVph} \simeq C_{\HII} \simeq
10$, where $C_{\HII}=\langle n_{\rm e}^2 \rangle/\langle n_{\rm e}
\rangle^2$ is the effective clumping factor of the ionised gas. Note
that the effective clumping, $C_{\HII}$, is the time averaged clumping
of the IGM from the time the first source turns on to the time of
reionisation. When the volume filling factor of the Str\"omgren
spheres is small the clumping factor of the ionised gas is very large,
but equals the clumping of the gas, $C$, at the redshift of
reionisation (see \fig~4 in paper~I). In addition, the
escape fraction of ionised photons, \fesc, and the value of $C_{\HII}$
are related to each other and depend on the resolution of the
hydrodynamic simulation or on the definition of \fesc in semianalytic
models.  Keeping in mind these caveats, the fraction of baryons that
needs to be converted into \pop3 stars to reionise the IGM is
approximately
\begin{equation}
\omega_* \simeq 4 \times 10^{-4} \left({C_{\HII} \over 10}\right)
\left({0.5 \over \langle f_{\rm esc} \rangle}\right) \left({2 \times
    10^{-3} \over \epsilon_{\rm UV}}\right) \left({X_{\rm UV} \over
    3}\right),
\label{eq:oms}
\end{equation}
where $X_{\rm UV}=\overline{h \nu} /(13.6 {\rm eV})$ is the mean
energy of UV photons in units of the \HI ionisation energy for a \pop3
spectrum.  By comparing \eq~\ref{eq:oms} to \eq~\ref{eq:omBH} it is
evident that the baryon fraction that needs to be accreted onto BHs to
partially reionise the IGM to \xe$=0.5$ is about 60 times smaller
than the fraction of baryons that needs to be converted into massive
\pop3 stars. This shows that X-rays from accreting BHs are potentially
more efficient than UV from stars in producing a large \taue.

Observations in the soft X-ray bands of Lyman-break
galaxies (which do not contain AGNs) at redshift $z \sim 3$
\citep{Nandra:02} have shown that their soft X-ray emission is
proportional to the UV emission: $L_{\rm X} /L_{\rm UV} \sim 0.01$.
This emission, attributed to X-ray binaries and SN remnants, is too
low to influence preionisation.  In addition, about 3\% of Lyman-break
galaxies contain central AGNs, accreting at roughly the Eddington
limit \citep{Steidel:02}. X-ray emission powered by accretion on
primordial seed BHs could thus be important. There are two scenarios
for the formation of supermassive black holes (SMBHs) in the bulges of
today's galaxies: monolithic collapse, or mergers and accretion on
seed BHs. Only the mass growth due to accretion produces X-rays
available to ionise the IGM. Assuming that most of the mass in SMBHs
observed in the bulges of galaxies at $z=0$ has been accreted at high
redshift, the upper limit for the accreted mass is
\begin{equation}
\omega_{\rm BH, max}= \left({M_{\rm SMBH} \over M_{\rm bulge}} \right)
\omega_{\rm bulge} = (1 \pm 0.7) \times 10^{-4}, 
\end{equation}
where we used $M_{\rm SMBH}/M_{\rm bulge}=(1.5 \pm 0.3)\times 10^{-3}$
\citep[\eg,][]{Kormendy:95, Gebhardt:00} and $\omega_{\rm bulge}=(6.5
\pm 3) \%$ \citep{Persic:92, Fukugita:98}.  But another scenario is
possible in which BH seeds grow through accretion at high redshift to
a level higher than currently observed, but a significant fraction of them is
expelled from the galaxies or does not end up in the observed
population of SMBHs. In this case a fraction of the halo DM would be made
of primordial BHs. The recent discovery of several ultra-luminous
X-ray sources (ULX) has been interpreted as evidence for
intermediate-mass BHs (about 100 M$_\odot$). These objects could also
be relics of primordial BHs.

In summary, X-rays may be more effective in increasing the optical
depth to electron Thompson scattering \taue with
respect to UV photons from stellar sources for the following reasons:
\bi
\item X-rays escape unabsorbed from the ISM of galaxies while for UV
 radiation \fesc$<1$.
\item X-rays ionise preferentially the low-density regions where the
  clumping factor is $C_{\HII}<1$. Consequently, the recombination rate
  is reduced.
\item X-rays only partially ionise the IGM: \xe$\sim 20-70$\%. This
  reduces the recombination rate by a factor \xe.
\item The emission efficiency of radiation due to gravitational
  accretion can be as high as $\epsilon \simeq 0.15$
  (accretion at the Eddington limit). This efficiency is about 100
  times the maximum efficiency produced by thermonuclear reactions in
  stars $\epsilon_{\rm UV}^{\rm max} \simeq 2 \times 10^{-3}$.
\item The accretion on seed BHs can continue even if the epoch of
  \pop3 star domination is very short due to the metal pollution of the
  star forming gas.  
\ei 
  On the other hand, X-rays may be inefficient with respect to UV
  radiation in ionising the IGM for the following reasons: 
\bi
\item X-rays have to be emitted at early times to be effective. This
  can only happen if a substantial density of seed BHs is formed in
  small-mass galaxies, at $z \simgt 20$.
\item X-rays are less efficient in ionising than UV photons. But, if
  the electron fraction in the IGM is \xe$ \simlt 10-20$ \%, secondary
  electrons are an important additional source of ionisation.
  Nevertheless, as we will show in \S~\ref{sec:res}, hard-UV photons
  and the redshifted X-ray background can enhance the IGM electron
  fraction to levels well above \xe$ \sim 10-20$\%.
\item X-rays are efficient in heating the IGM and therefore they
  increase the IGM Jeans mass. This will have a negative feedback on
  the formation of the smaller mass galaxies and consequently on the
  formation of seed BHs and their ability to accrete gas.  
\ei

\section{Mass Accretion History on seed BHs}\label{sec:xsour}

\begin{figure*}
\centerline{\psfig{figure=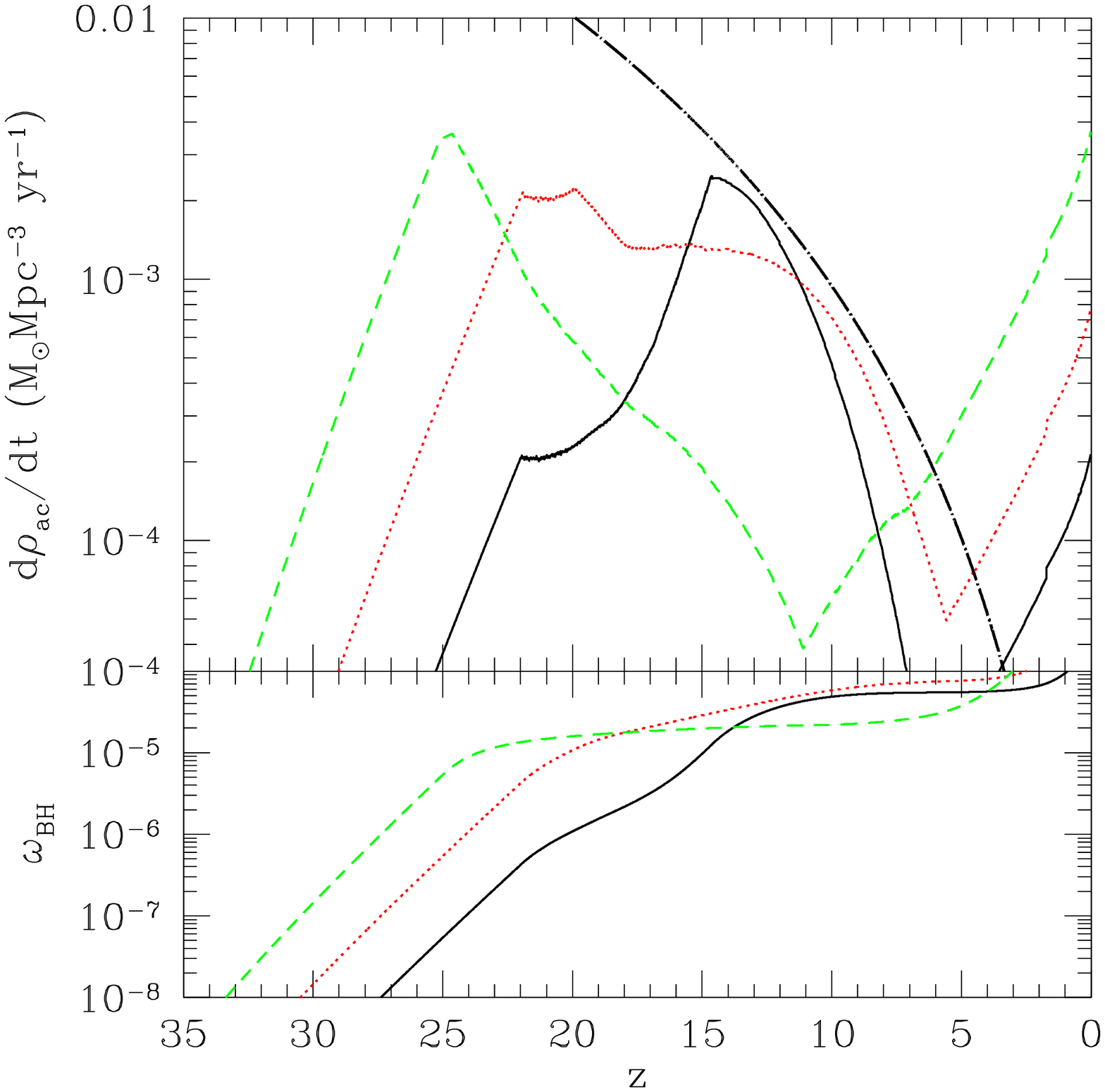,width=9cm}
\psfig{figure=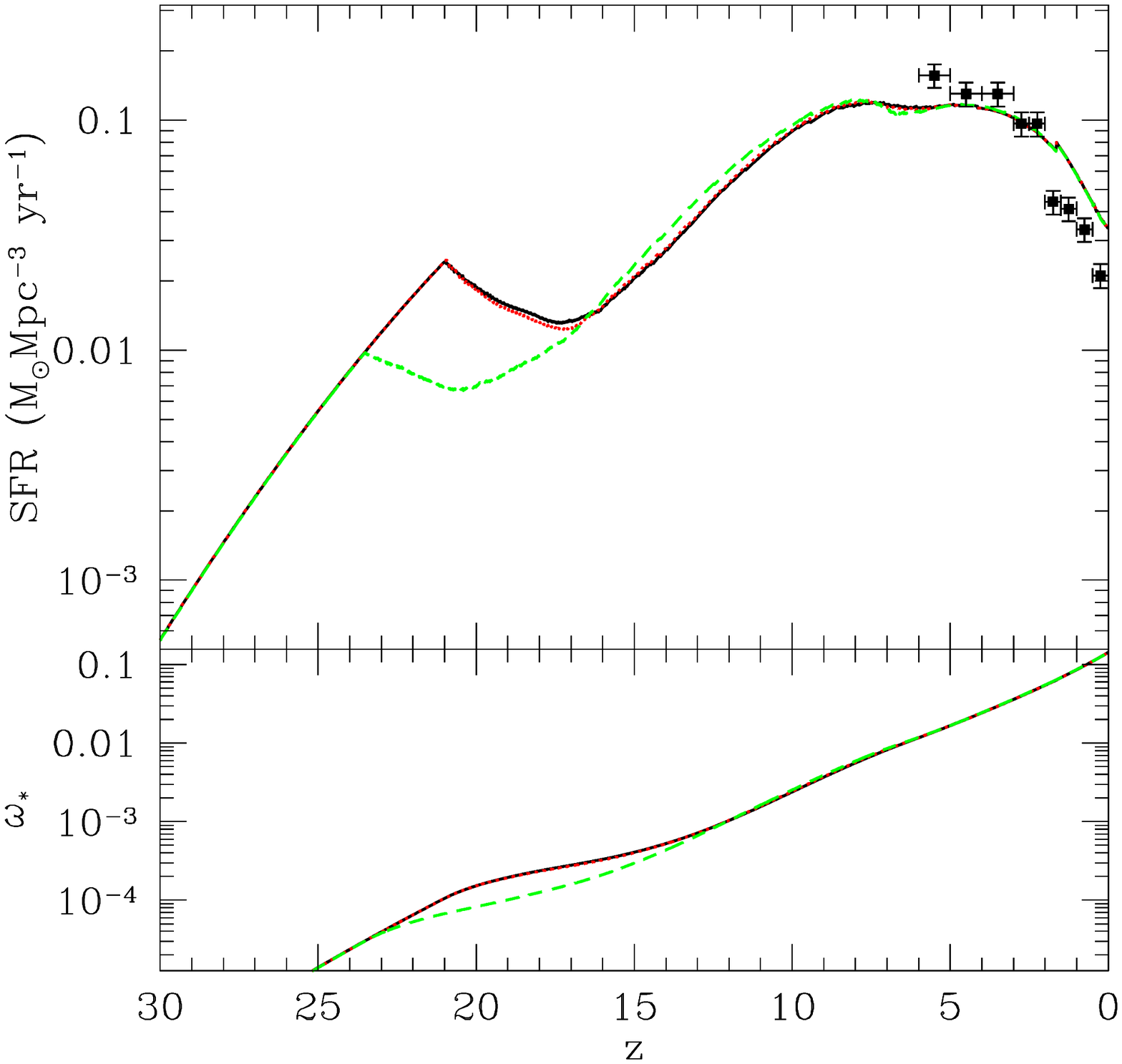,width=9cm}}
\caption{\label{fig:sfr} (left) The top panel shows the black hole
  accretion rate, $\dot \rho_{\rm ac} \equiv \rho_0 \dot \omega_{\rm
    ac}$ (where $\rho_0$ is the cosmic baryon density at $z=0$ and
  $\omega_{\rm ac}$ is given in \eq~[\ref{eq:omac}]), as a function of
  redshift. The bottom panel shows the baryon fraction in BHs,
  $\omega_{\rm BH}$, as a function of redshift. (right) The top panel
  shows the star formation rate, $\dot \rho_{*} \equiv \rho_0 \dot
  \omega_{*}$, as a function of redshift. The bottom panel shows the
  baryon fraction in stars as a function of redshift. The solid,
  dotted and dashed lines refer models M3, M4, and M5 in
  \tab~\ref{tab:1}, respectively. The dot-dashed line in the top panel
  of the left figure is an upper limit for the accretion rate given by
  \eq~(\ref{eq:maxac}).}
\end{figure*}

The formation rate of seed black holes from star formation is very
uncertain. 
The calculations of the mass of the remnant and metal yields are
strongly dependent on the energy of the SN explosion,
$E_{51}=E/10^{51}$ erg, on the mass shell at which the explosion
takes place and the star metallicity. These numbers are uncertain and
calculations \citep{WoosleyW:95} have been done for a grid of models.
In \tab~\ref{tab:rem} we list the fraction, $f_{\rm BH}$, of stellar
mass expected to end in seed BHs for the different explosion models,
gas metallicities and stellar IMF. The data is from
\cite{WoosleyW:95}: Model A has $E_{51} \sim 1$; Model B has $E_{51}
\sim 2$; and Model C has $E_{51} \sim 3$.
The values listed in the table are the weighted means of $f_{\rm BH}$
for stellar masses between $12~{\rm M}_\odot<M_*<40$ M$_\odot$ with
weight $M^{\alpha}$.  Note that for a Salpeter IMF ($\alpha=-1.35$)
with stellar masses between $1~{\rm M}_\odot<M_*<100$ M$_\odot$, the
mass in stars with $M_*>8$ M$_{\odot}$ is 20 \% of the total.

Very massive stars (VMS) in the mass range $140~{\rm M}_\odot<M_*<260$
M$_\odot$ are believed to explode completely without leaving any
remnant due to an instability produced by annihilation of electrons
and positrons (pair-instability SNe) and hence have $f_{\rm BH}=0$.
Stars more massive than $M_* = 260$ M$_\odot$ are believed to collapse
directly into BHs, without exploding as SNe and hence have $f_{\rm
  BH}=100$\% \citep{HegerW:02}.  The existence of a mixture of masses
for supermassive \pop3 stars allows for all possible values of $f_{\rm
  BH}$ from zero to unity.
\begin{table}
\centering
\caption{Fraction of stellar mass that forms seed BHs}\label{tab:rem}
\begin{tabular*}{6.5 cm}[]{c|ccc|c}
$Z/Z_\odot$ & & $f_{\rm BH}$ (\%) & & IMF\\
            & Model A & Model B & Model C & $\alpha$ \\
\hline
$0$ & 19 & 12 & 11 & -1.35\\
$0$ & 24 & 12 & 8 & 0\\
$0$ & 24 & 13 & 7 & 1\\

$10^{-4}$ & 10 & 8 & 8 & -1.35\\
$10^{-4}$ & 13 & 8 & 7 & 0\\
$10^{-4}$ & 16 & 8 & 7 & 1\\

$1$ & 11 & 10 & 8 & -1.35\\
$1$ & 13 & 10 & 8 & 0\\
$1$ & 15 & 10 & 8 & 1\\
\end{tabular*}
\end{table}
 


The mass growth of seed BHs can be calculated if we know the accretion
rate $\dot \omega_{\rm ac}$ of BHs, the formation rate of seed BHs, $\dot
\omega_{\rm seed}$, and the ejection rate of BHs from galaxy haloes, $\dot
\omega_{\rm ej}$. To calculate the mass growth of BHs we solve the
following equation:
\begin{equation}
\dot \omega_{\rm BH} = \dot \omega_{\rm ac} -\dot \omega_{\rm ej} +
\dot \omega_{\rm seed},
\label{eq:rate}
\end{equation}
where $\omega_{\rm BH}, \omega_{\rm ac}, \omega_{\rm ej}$ and
$\omega_{\rm seed}$ are expressed as a fraction of the baryon content
of the universe. The accretion rate is proportional to $\omega_{\rm
  BH}$, independently of the mass function of BHs,
\begin{equation}
\dot \omega_{\rm ac} = f_{\rm duty}{\omega_{\rm BH} \over t_{\rm Ed}}, 
\label{eq:omac}
\end{equation}
where the accretion time, $t_{\rm ac} \equiv t_{\rm Ed}/f_{\rm duty}$,
is expressed as the fraction of time, $f_{\rm duty}$, that the BHs are
accreting at the Eddington rate. If we assume that the BH accretes at
the Eddington limit for a time interval $t_{on}$ and stops accreting
for a time interval $t_{\rm off}$, the duty cycle parameter is $f_{\rm
  duty} = t_{\rm on}/(t_{\rm off}+t_{\rm on})<1$.  If we neglect BH
ejection we have
\begin{equation}
\dot \omega_{\rm BH} = f_{\rm duty}{\omega_{\rm BH} \over t_{\rm Ed}} + f_{\rm BH}\dot \omega_*,
\label{eq:ombh}
\end{equation}
where we have assumed that the formation rate of seed BHs is a
fraction, $f_{\rm BH}$, of the star formation rate $\dot \omega_{\rm
  seed}=f_{\rm BH} \dot \omega_*$ (see \tab~\ref{tab:1}).  Integrating
\eq~(\ref{eq:ombh}) we get
\begin{equation}
\omega_{\rm BH}(t)=\int_0^t dt^\prime f_{\rm BH} \dot \omega_{*}(
t^\prime)\exp{[F(t)-F(t^\prime)]},
\label{eq:ombh1}
\end{equation}
where,
\[
F(t)={1 \over t_{\rm Ed}}\int_0^t f_{\rm duty} dt^\prime.
\]
We parameterise the duty cycle as
\begin{equation}
f_{\rm duty}(z)=\left({1+z \over z_{\rm BH}}\right)^\phi,
\label{eq:fduty}
\end{equation}
and imposing the condition $0.001 \le f_{\rm duty} \le 1$ to be
consistent with the observed fraction of AGN $f_{\rm duty} \simlt 3$\% at $z
\sim 3$ \citep{Steidel:02}, and $f_{\rm duty} \sim 0.001$ at $z=0$. 

In this work we will not attempt to investigate in detail the physical
processes and feedback effects that could produce the duty cycle given
in \eq~(\ref{eq:fduty}).  Here we simply point out the existence of,
at least, one simple physical model that can reproduce the family of
equations given in (\ref{eq:fduty}). In this model the time $t_{\rm
  off}$ during which the BH is quiescent is the Compton cooling time
$t_{\rm Comp}=(224~{\rm Myr})[(z+1)/10]^{-4}$.  This is the time that
it takes for the gas in the proximity of the BH to cool and be
accreted after the temporary halt of the accretion produced by gas
heating from the previous BH activity. If we further assume that the
duration of the burst is a fraction, $\eta$, of the Eddington time
($t_{\rm on}=\eta t_{\rm Ed}$) we can reproduce $f_{\rm duty}\equiv
1/(1+t_{\rm off}/t_{\rm on})$ with the form given in
\eq~(\ref{eq:fduty}), where the value of $z_{\rm BH}$ is determined by
the value of $\eta$.  For reasonable values of $\eta$ in the range
$1-100$\% it is possible to describe models spanning from the early
(\ie, adopting $\eta \sim 1$\%) to the late preionisation (\ie,
adopting $\eta \sim 100$\%). On the basis of observations of AGNs at
redshift $z \simlt 3$, the best estimate of $t_{\rm on}$ is $\sim 10$ Myr
\cite[\eg,][]{Haehnelt:98}.  This suggests that the intermediate
preionisation models (that adopt $\eta \sim 10$\%) are prefered.

For the purpose of this paper we do not need to focus on the
functional form of the duty cycle at low redshifts ($z \simlt 3$)
provided that we can avoid to violate observational constraints on the
duty cycle and BH accretion rate. In order to get a realistic
accretion rate at lower redshift in some models, it might be necessary
to adopt a non zero ejection rate of BHs from the host galaxies [\ie,
$\dot \omega_{\rm ej}\not= 0$ in \eq~(\ref{eq:rate})]. Also this
assumption has physical motivations \citep[\eg,][]{MadauR:03}.

In \S~\ref{sec:res_sa} we use a semianalytic code to follow the
chemical and thermal history of the IGM in models where accretion on
seed BHs partially ionise the universe. At redshift $z \approx 6.3$
stellar sources reionise the IGM to a level consistent with Sloan
quasar observations. We explore three cases that differ for the
accretion history onto seed BHs. The accretion history, parametrised
using the equations derived in this section, can happen early ($z \sim
25$), late ($z \sim 15$) or at intermediate redshifts ($z \sim 20$).
If we wished to allow for a lag, $\Delta t$, between the formation of
an individual BH and the commencement of accretion on it
\citep[\cf,][]{MadauR:03,WhalenAN:03} we would simply replace $\dot
\omega_*(t^\prime)$ with $\dot \omega_*(t^\prime-\Delta t)$ in
\eq~(\ref{eq:ombh1}). In \tab~\ref{tab:1} we list the parameters
adopted for the semianalytic simulations whose results will be shown
in \S~\ref{sec:res_sa}. The values for $f_{\rm BH}$ allow for a range
of mass functions from Salpeter to top-heavy. We also set $f_{\rm
  BH}=0$ in all models after $z=15$ since the formation of \pop3 stars
that produce the seed BHs is probably self-terminated by metal
enrichment well before reionisation at $z_{\rm rei}=7$. But this
assumption does not have any effect on the BH accretion history
because the total mass in BHs becomes dominated by the accretion on
e-folding time scales $t \sim t_{\rm ac}$, making the contribution by
further seed BH formation negligible after the initial phase (before
the global accretion rate reaches the maximum).
\begin{table}
\caption{Parameter of the semianalytic models}\label{tab:1}
\begin{tabular*}{7.5 cm}[]{l|cclcc}
Model & $f_{\rm BH}$ (\%)& $\phi$ & $z_{\rm PBH}$ ($t_{\rm on}$ Myr) & $n_{\rm s}$ &
$z_{\rm PopIII}$\\
\hline
M3 & 0.2 & 12 & 15 ~($\sim 100$) & 1.00 & -\\
M3b & 0.1 & 8 & 15 ~($\sim 100$) & 1.04 & -\\
M4 & 2 & 8 & 20 ~($\sim 10$) & 1.00 & -\\
M4b & 2 & 8 & 20 ~($\sim 10$) & 1.00 & 15\\
M5 & 20 & 12 & 25 ~($\sim 1$) & 1.00 &  -\\
\end{tabular*}
\\

Meaning of the parameters:
$f_{\rm BH}$ is the fraction of stellar mass that collapses into BHs;
$\phi$ and $z_{\rm PBH}$ are the parameters in \eq~\ref{eq:fduty} that 
determines the time evolution of the duty cycle $f_{\rm duty}$;  
$n_{\rm s}$ is the spectral index of the spectrum of primordial
density perturbations; $z_{\rm PopIII}$ is the redshift at which UV
radiation from \pop3 stars (that has emissivity 11 times larger than
the UV from \popII stars) begins to fade: the function for the UV
emissivity is given by 
$\epsilon_{\rm UV}(z)/\epsilon_{\rm UV}(z=0) =10
\{\arctan{[2(z-z_{\rm PopIII})]}/\pi+0.5\}+1$, where the
emissivity from \popII stars (that is used in all the other models) is $\epsilon_{\rm UV}(z=0)=4.8 \times 10^{-5}$. 
\end{table}

The BH accretion rate as a function of redshift, $\dot \rho_{\rm
  ac}(z) \equiv \rho_0 \dot \omega_{\rm ac}$, where $\rho_0$ is the
cosmic baryon density at $z=0$ and $\omega_{\rm ac}$ is given in
\eq~(\ref{eq:omac}), is shown in the top panel of
\fig~\ref{fig:sfr}(left). The baryon fraction in BHs, $\omega_{\rm
  BH}$, as a function of redshift is shown in the bottom panel of
\fig~\ref{fig:sfr}(left).  The solid, dotted and dashed lines refer
models M3, M4, and M5 in \tab~\ref{tab:1}, respectively. As already
noted, all the models have about the same $\omega_{\rm BH} \sim
10^{-4}$ at $z=0$ (\ie, the same total energy output from the sources)
and differ for the characteristic epoch of accretion. The BH accretion
rate is always negligible at the redshift, $z_{\rm rei}$, of IGM
reionisation by stellar sources.  The dot-dashed line shown in the top
panel of the \fig~\ref{fig:sfr}(left) is an upper limit for the
accretion rate.  This limit on the accretion rate is given by
\eq~(\ref{eq:maxac}), that will be discussed in \S~\ref{sec:res}.

In \fig~\ref{fig:sfr}(right) we show the star formation rate (top
panel) and the baryon fraction in stars, $\omega_{*}$ (bottom panel),
as a function of redshift for the same models (\cf, \tab~\ref{tab:1}).
The points with errorbars in the top panel show the observed SFR from
\citep{Lanzetta:02}. The baryon fraction in stars at redshift $z=0$ is
$\omega_*=20$\% in all the models, consistent with observations
\citep{Persic:92}.  The feedback due to the X-ray heating of the IGM
suppresses galaxy formation in haloes with masses smaller than the IGM
filtering mass, producing the substantial reduction of the global SFR
observed at redshifts around $z=15$ in the top panel of
\fig~\ref{fig:sfr}(right).
\begin{figure}
\centerline{\psfig{figure=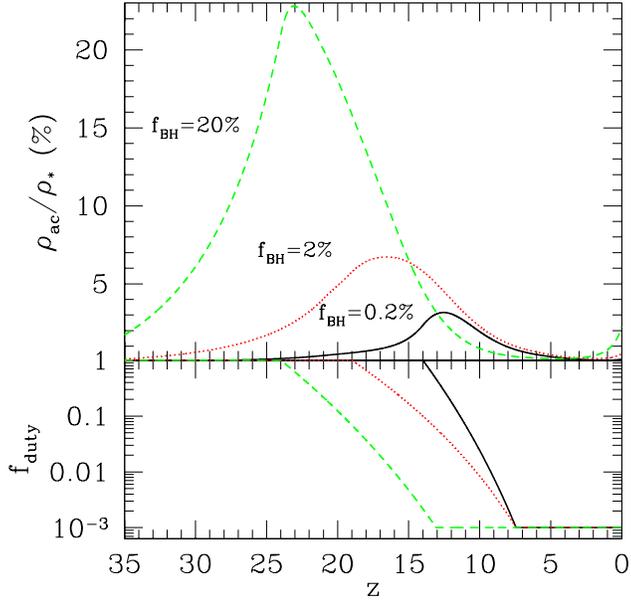,width=8.5cm}}
\caption{\label{fig:fBHM} The top panel shows the mass fraction
  accreted onto seed BHs with respect to the star fraction in the
  universe. The bottom panel shows the duty cycle, $f_{\rm duty}$,
  assumed in each model. The solid, dotted and dashed lines refer
  models M3, M4, and M5 in \tab~\ref{tab:1}, respectively.}
\end{figure}

Finally, in \fig~\ref{fig:fBHM}, we show the mass fraction accreted
onto seed BHs with respect to the star fraction in the universe,
$\rho_{\rm ac}/\rho_*=\omega_{\rm ac}/\omega_*$ (top panel), and the
adopted duty cycle (bottom panel). Again, the dashed, dotted and solid
lines refer to models M3, M4 and M5 in \tab~\ref{tab:1}. As shown by
the labels next to each line, the early accretion model has
$f_{\rm BH}=20$\%, therefore it requires that most of the mass of \pop3
stars implodes into BHs. But for the other models only a fraction
$f_{\rm BH} \simlt 2$\% of the stars needs to implode into BHs. This
condition does not require a top-heavy IMF.  It is interesting to note
(but not surprising) that in order to have a substantial mass
accretion onto seed BHs at early times a large fraction of seed BHs,
$f_{\rm BH}$, from \pop3 stars is required. This is because the accreted
mass increases exponentially with e-folding time scale $t_{\rm Ed} \simeq
10^{8}$ yrs, which is about the Hubble time at $z=30$. To conclude we
note that if seed BHs are not able to accrete initially at the
Eddington rate (see bottom panel of \fig~\ref{fig:fBHM} that shows the
adopted duty cycle), a larger fraction of seed BHs, $f_{\rm BH}$, is
necessary in order to get the same global accretion rate as in our
models. This is clearly possible to do for the intermediate and late
accretion models, but would need extremely high and unrealistic values
of $f_{\rm BH}$ in the early accretion model M3.


\subsection{Spectral Energy Distribution}\label{ssec:sed}

\begin{figure}
\centerline{\psfig{figure=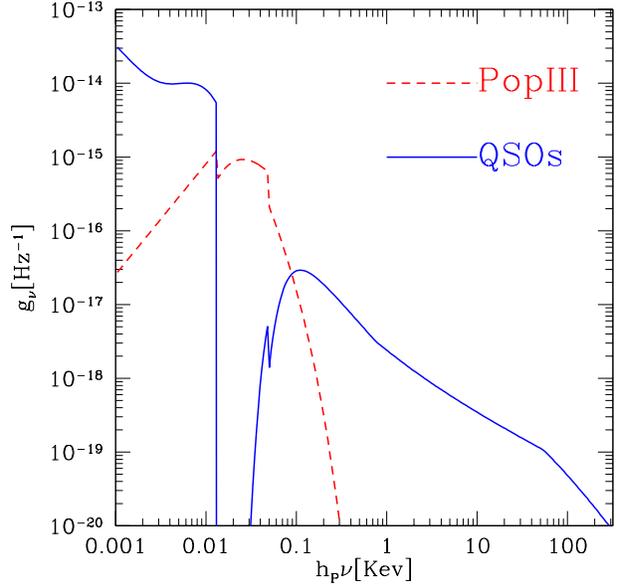,width=8.5cm}}
\caption{\label{fig:sed} Spectral energy distribution of \pop3
  stars (dashed line) and quasars (solid line). UV radiation is mainly
  emitted by stellar sources and X-rays by quasars. We assume that
  most UV radiation from quasars does not escape to the IGM but it is
  absorbed by an obscuring torus or in the ISM. In the adopted quasar
  SED the thermal component, that depends mildly on the BH mass, is
  negligible with respect to the power low component that is mass
  independent.}
\end{figure}

In \fig~\ref{fig:sed} we show the spectral energy distribution (SED)
of stars and quasars adopted in our simulations. For the stars we
adopt a SED calculated for zero-metallicity stars \citep{Tumlinson:00}
and for quasars we use a spectrum similar (but not identical) to the
one calculated by \cite{SazonovO:04}. Their spectrum is a template,
inferred on the basis of observations, representing the average quasar
in the universe and determined by considerations both of the (X,
$\gamma$) ray background and the spectra of individual objects. The
quasar spectrum has an X-ray bump produced by absorption of UV photons
by intervening gas and an IR excess produced by the reprocessed UV
photons. Using this SED we can distinguish between reionisation by UV
photons produced mainly by the stars and X-rays produced by quasars.
The energy band that contributes the most to reionisation, also
amplified by the effect of secondary electrons, is 0.5 keV $< h_p \nu
< 5$ keV.  But as we will point out in paper~IIb, the X-ray background
photons that are redshifted to the extreme ultraviolet (EUV) bands can
be the dominant source for preionisation in the voids.

The typical expected mass of the accreting BHs in our model is $M_{\rm
  BH} \simlt 100-1000$ M$_\odot$ at the high redshifts of interest,
while the adopted SED is inferred from the accretion onto BHs with
masses of $M_{\rm BH} \simlt 10^{5}-10^6$ M$_\odot$. Generally
speaking the SED of accreting compact objects has two components that
contribute comparably to the total energy output: (i) a power law
component from the hot corona around the BH and (ii) a thermal
component from the inner part of the hot accretion disk. The power law
component is independent of the mass of the accreting object (from
X-ray binaries to AGN) while the disk component has a temperature that
depends mildly on the mass of the BH as $kT_{\rm disk} \approx
1.2~{\rm keV} (M_{\rm BH}/ 10~{\rm M}_\odot)^{-1/4}$. The observed
spectra of ULX sources tentatively show a subdominant soft component
fitted by multicolour disks blackbody with $kT_{\rm disk} \sim 0.15$
keV, which has been interpreted as a tentative evidence of
intermediate mass black holes \citep{Miller:03}. The non-thermal
component, which dominates the total energy output, is a power law
similar to the one adopted in the present paper.  Note that the
spectrum that we have assumed is also similar to the
\cite{SazonovO:04} ``absorbed quasar spectrum'' (\fig~\ref{fig:sed})
were the UV bump produced by the thermal component in supermassive BHs
is absorbed and in fact has no UV at all. The spectrum that we adopt
is therefore dominated by the power law component (absorbed in the UV)
that is roughtly independent of the assumed mass of the accreting BH.

\section{Semianalytic Models}\label{sec:res_sa}

\begin{figure*}
\centerline{\psfig{figure=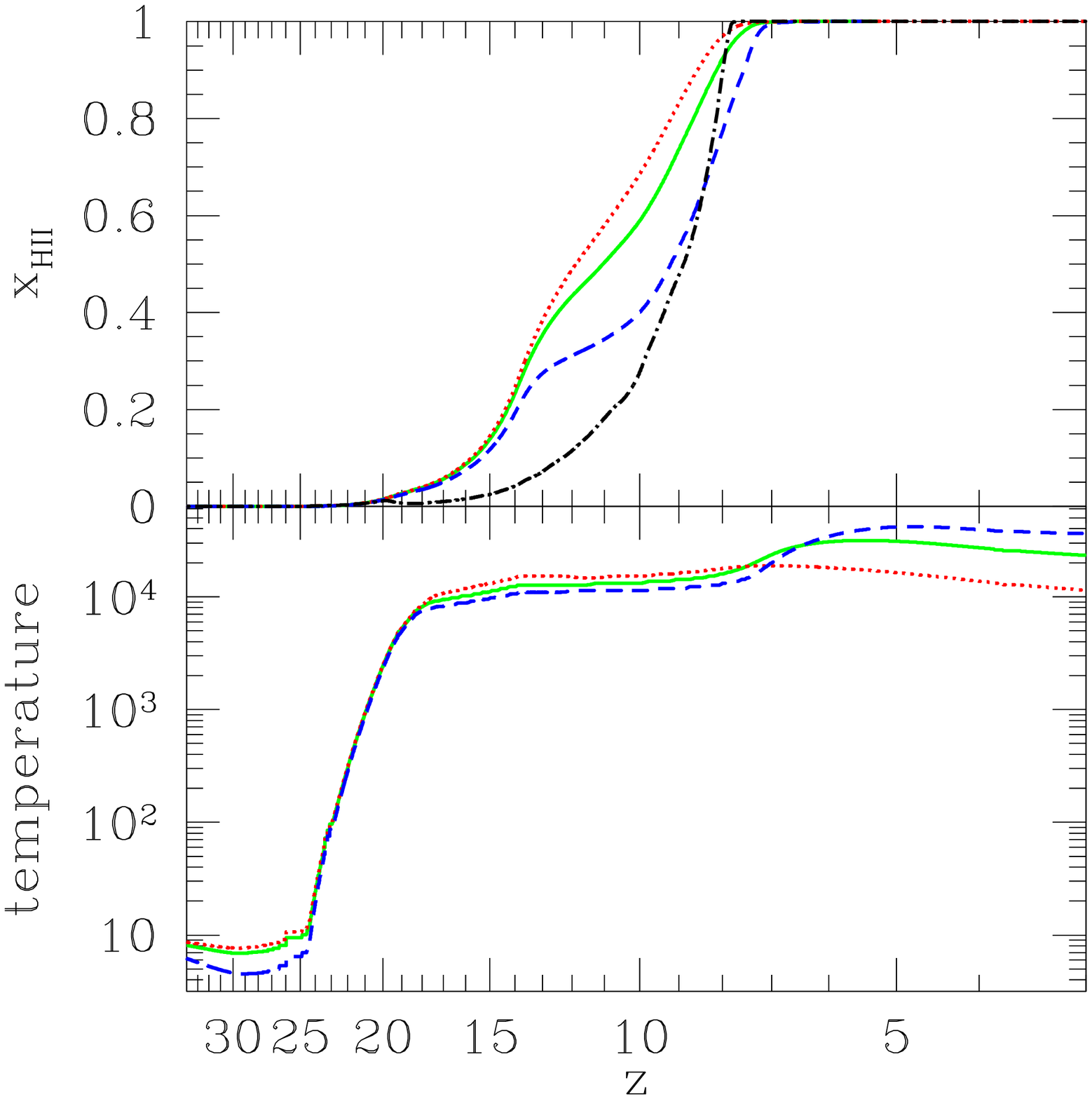,width=6cm}
\psfig{figure=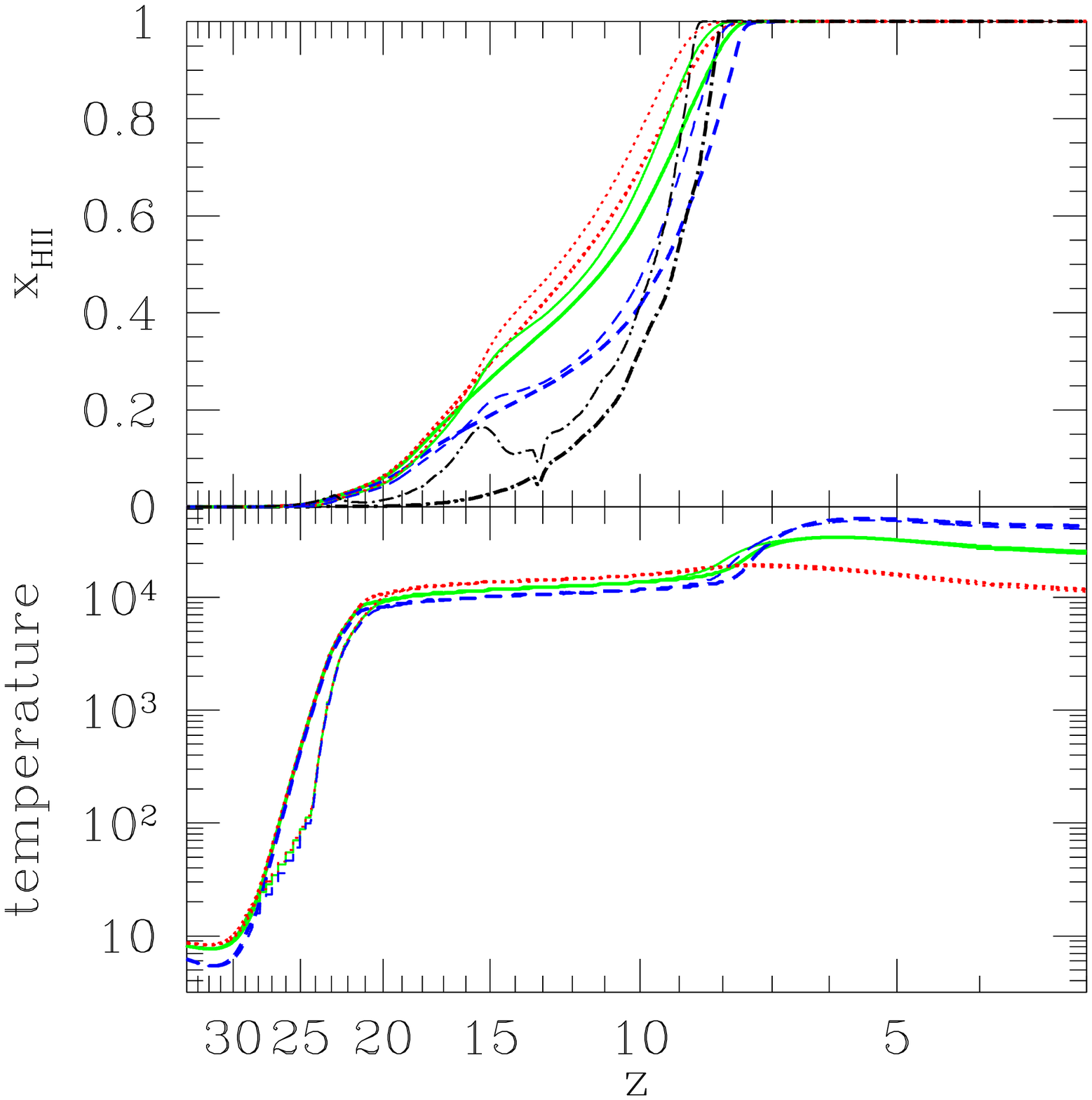,width=6cm}
\psfig{figure=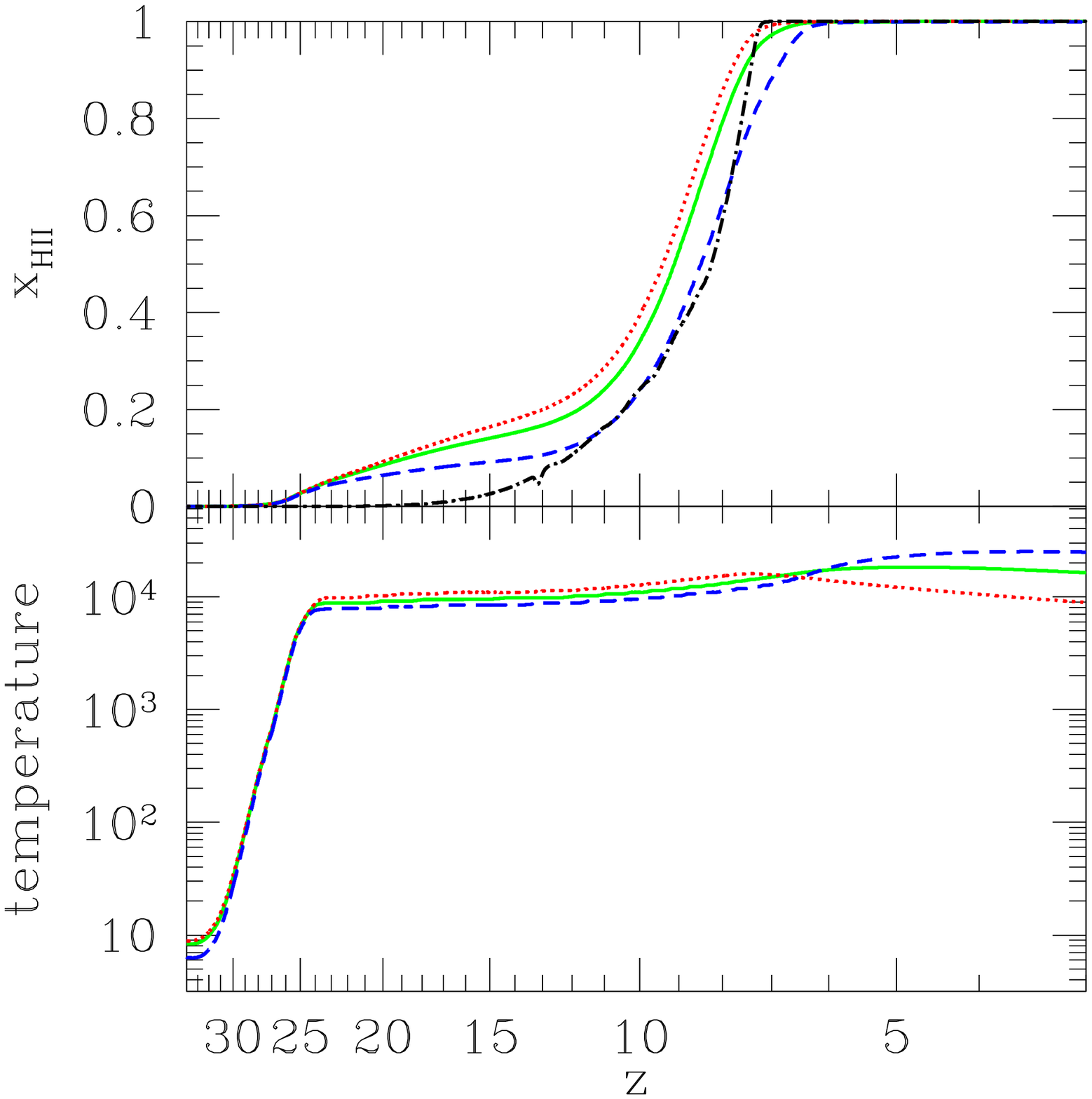,width=6cm}}
\caption{\label{fig:mod1} (left) The top panel shows the ionised
  hydrogen fraction and the volume filling factor of \HII regions
  (dot-dashed line) as a function of redshift for model M3. The bottom
  panel shows the IGM temperature as a function of redshift for model
  M3.  The dotted, solid and dashed lines refer to gas over-densities
  $\delta=0.1, 1, 10$ in the IGM outside \HII regions, respectively.
  After overlap, when the ionising background is uniform, the lines
  refer to the ionised IGM. (centre) Same as in the left figure but
  for model M4 (thick lines) and M4b (thin lines). (right) Same as in
  the left figure but for model M5.}
\end{figure*}

In this section we show the results found using a semianalytic model
for reionisation. The advantage of the semianalytic approach with
respect to the cosmological simulations presented in paper~IIb is that
we can explore a larger parameter space and the results are not
affected by the resolution of the simulations.  We use this approach
to study the dependence of the Thomson optical depth on cosmological
parameters and accretion histories of seed BHs.  The results presented
in this section also helped to derive a physically motivated, time
dependent X-ray emissivity, to be used in the more computationally
expensive numerical simulations.  In addition, the semianalytic models
are an aid for the interpretation of the results of cosmological
simulations, made complicated by the interplay of a larger number of
physical processes.  If not otherwise specified, we adopt the
concordance $\Lambda$CDM cosmological model with $h=0.72$,
$\Omega_{\rm m}=0.27$, $\Omega_{\rm b}=0.046$, $n_{\rm s}=1$ and
$\sigma_8=0.91$, that is consistent with the WMAP measurements
($h=0.71^{+0.04}_{-0.03}$, $\Omega_{\rm
  m}h^2=0.135^{+0.008}_{-0.009}$, $\Omega_{\rm b}h^2=0.0224 \pm
0.0009$, $n_{\rm s}=0.93 \pm 0.03$ and $\sigma_8=0.84 \pm 0.04$,
\citealp{Spergel:03}).

\subsection{The code}
We implemented a semianalytic model to study IGM reionisation,
chemical evolution and re-heating. The filling factor of \HII regions
is calculated following the method in \cite{Chiu:00}. But we also
include X-ray ionisation of the IGM before overlap of the \HII
regions.  A detailed description of the code is given in
Appendix~\ref{ap:A}. Here we summarise the main processes included in
the code at this stage: 
\bi
\item We calculate the mass function of DM haloes and the formation
  rate of haloes using the extended Press-Schechter formalism.  The
  star formation rate (SFR) is proportional to the formation rate of
  haloes. In the integral for the global SFR we take into account
  cooling and dynamical biases that depend on the mass of the haloes.
\item We include thermal feedback on star formation: the minimum mass of
  star-forming galaxies at a given redshift is determined by the IGM
  filtering mass.
\item The emissivity of accreting BHs is calculated as in
  \S~\ref{sec:xsour}.
\item We assume a two-phase IGM. One phase is the fully ionised gas
  inside the \HII regions and the other is the partially-ionised or
  neutral gas outside the \HII regions.
\item Radiative transfer is solved by splitting the spectrum into \HI
  optically thick radiation and background radiation. Given the
  clumping factor inside \HII regions we follow the evolution of
  the filling factor, temperature and chemistry of the ionised gas.
  Outside the \HII regions the temperature and element abundances are
  calculated as a function of the overdensity.
\item We calculate the time-dependent chemical network for 8 ions (H,
  H$^+$, H$^-$, H$_2$, H$_2^+$, He, He$^+$, He$^{++}$) and thermal
  evolution inside and outside \HII regions. The following heating and
  cooling processes are included: collisional and photo-ionisation,
  ionisation by secondary electrons, H, He and \H2 cooling, Compton
  cooling-heating and the cosmological expansion term. We neglect
  cooling from metal contamination in small-mass galaxies. The rates
  are from \cite{RicottiGS:01}.  
\ei

\subsection{Results}\label{sec:res}

We have run several models to study the sensitivity of the results to
the accretion history on seed BHs and to cosmological parameters. In
\tab~\ref{tab:1} we list the parameters of the models shown in the
next figures. In all the models the integrated energy emitted by
accreting BHs is the same and stellar reionisation happens at $z\sim
7$.  We normalise the BH accretion rate assuming that the total mass
in BHs at $z=0$ is $\omega_{\rm BH, max}=10^{-4}$, the mass of SMBHs in
the galactic centres. This assumption is rather conservative since it
is likely that a fraction of seed BHs is not incorporated into the
SMBH population at $z=0$. During galaxy mergers there is a
non-negligible probability that one of the two merging BHs sitting in
the bulge is expelled from the galaxy \citep[\cf,][]{MadauR:03}. This
probability is larger for unequal mass BHs, and happens in the final
phase of the merger during the phase where gravitational radiation is
emitted.

In summary the constraints on the semianalytic models are the following:
\begin{itemize}
\item The integrated energy output from accreting BHs is limited by
  assuming that the total mass in BHs at $z=0$ equals the total mass
  of SMBHs in the observed in the centres of galaxies. This
  requirement is a very conservative one, since theoretical
  calculations \citep[\eg,][]{MadauR:01} show that most intermediate
  mass BHs formed at high redshift will not merge into the SMBHs in
  the galactic centres.
\item Seed BHs accrete initially at the Eddington rate. The duty cycle
  in the late preionisation model (M3) is consistent with observations
  of QSO at $z \sim 3$ that show a duty cycle of 3\% and with QSO at
  $z \sim 0$ that show a duty cycle of 0.1\%
\item The star formation efficiency is constrained by fitting the
  observed star formation rate (Madau plot) at $z<6$
  \citep{Lanzetta:02}. The integrated baryon fraction in stars at $z=0$
  is consistent with the one estimated by \cite{Fukugita:98}.
\item The effective UV emissivity from stellar sources is constrained
  to reproduce the transmitted flux observed in the Sloan quasars at
  $z<6.2$. Assuming a Salpeter IMF the adopted escape fraction is
  \fesc$=30$ \%.
\item The specific intensity of the ionising background at $z<6$ is
  consistent with observations.
\end{itemize}

\begin{figure}
\centerline{\psfig{figure=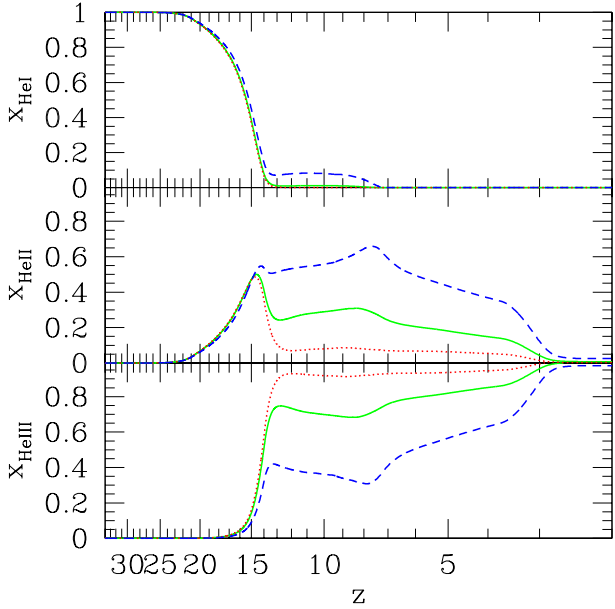,width=8.5cm}}
\caption{\label{fig:helM3}  Helium ionisation history for model M3. The
  dotted, solid and dashed lines refer to gas over-densities
  $\delta=0.1, 1$, and 10, respectively.}
\end{figure}

The typical ionisation and thermal history of the X-ray preionisation
scenario is illustrated in \fig~\ref{fig:mod1} for hydrogen and
\fig~\ref{fig:helM3} for helium.  In \fig~\ref{fig:mod1} we show the
hydrogen ionisation (top panel) and the IGM thermal history (bottom
panel) for model M3 (figure on the left), model M4 and M4b (figure in
the centre) and for model M5 (figure on the right). In model M3, BH
accretion takes place mostly at $z>15$, at $z>20$ in model M4, and at
$z>25$ in model M5.  The dotted, solid and dashed lines refer to gas
over-densities $\delta=0.1, 1$, and 10, respectively. Before
reionisation at \zrei$\sim 7$, the lines refer to the gas outside the
\HII regions.  After overlap, when the ionising background is uniform,
the lines refer to the fully ionised IGM. The dot-dashed lines in the
top panels show the volume filling factor of the \HII regions as a
function of redshift. In the figure at the centre the thin lines show
a model (M4b) where the UV emissivity is about 10 times larger at $z>
15$ than at $z<15$, to account for the contribution of a top-heavy
\pop3 star population. The volume filling factor of the \HII regions
in this model differs from the same model where the \pop3 contribution
is neglected (M4). The optical depth to Thompson scattering in model
M4 and M4b has about the same value. 

As even the earliest investigations showed \citep{OstrikerG:96},
reheating significantly precedes reionisation (weather or not X-ray
ionisation is included).  
In any X-ray preionisation model, by the time the ionisation fraction
in the IGM reaches \xe$ \sim 10$\%, the IGM has a temperature $T
\approx 10^4$ K (\cf, \fig~\ref{fig:mod1}). This happens in all the
models independently of the starting redshift of preionisation because
the ionising efficiency of secondary electrons is reduced when \xe$
\simlt 10$\% and the energy of X-ray photons is mostly converted into
heat.

By inspecting the dotted, solid, and dashed lines in the bottom panels
of \fig~\ref{fig:mod1}, that show the gas temperature at overdensities
$\delta=0.1, 1$, and 10, respectively, we identify three epochs in the
reheating history of the IGM.  Initially, before complete reionisation
by UV radiation, the temperature of the IGM outside the cosmological
\HII regions is almost isothermal ($T_{\rm IGM} \approx 10^4$ K) with
overdense regions slightly cooler than underdense regions
\footnote{This is due to the shorter cooling time at higher densities.
  At $T_{\rm IGM} \approx 10^4$ K, \H2 cooling is dominant over the
  \lya cooling. \H2 cooling is not very important and the dependence
  of the temperature on the density is weak.}.  At reionisation the
IGM becomes isothermal.  After reionisation the IGM temperature
dependence on the density follows a tight relationship, often referred
to as ``the effective equation of state of the IGM'', $T_{\rm IGM}
=T_{\rm IGM, 0} (1+\delta)^{(\gamma-1)}$, where $\delta$ is the
overdensity \citep{Hui:97}.  The parameter $(\gamma-1)$ is zero (\ie,
the IGM is isothermal) at reionisation and $(\gamma-1)>1$ (\ie,
overdense regions are hotter than the mean density gas) after
reionisation.  The temperature $T_{\rm IGM,0}$ at reionisation is
determined by the spectrum of the background radiation: it is higher
the harder the spectrum. If the spectrum does not evolve afterwards,
the temperature of the IGM is expected to decrease almost
adiabatically (\eg, $(\gamma-1) \sim 0.6$).

In our models the adopted spectrum at $z_{\rm rei}=7$ is consistent
with the temperature and $(\gamma-1)$ at $z=3-5$ derived from the line
widths of the \lya forest: $T_{\rm IGM, 0} \sim 1-3 \times 10^4$ K,
$(\gamma-1) \sim 0.1-0.4$ \citep{RicottiGS:00, Schaye:00, Theuns:02}.
We find that the IGM temperature decreases slowly with redshift
($T_{\rm IGM} \propto (1+z)^{0.5}$) and the equation of state has
$(\gamma-1) =0.3 \pm 0.1$ (\ie, an intermediate value between the
isothermal and adiabatic case). Loose constraints (due to the large
errors on the measured temperature) can be put on reionisation models
using the observed $T_{\rm IGM}$. If reionisation happens too early
and the IGM is not reheated by additional sources that produce the
hardening of the background spectrum, the temperature at $z \sim 3-5$
would be too low when compared to observations. Or, if the spectrum is
extremely hard at $z_{\rm rei} \simeq 7$, the IGM temperature at $z
\simeq 3-5$ would be higher than observed.  Compton heating is always
much less efficient than photoionisation heating in these models. We
have estimated that it is subdominant if the neutral fraction is
$x_{\HI} \simgt 10^{-7}$.

\begin{figure}
\centerline{\psfig{figure=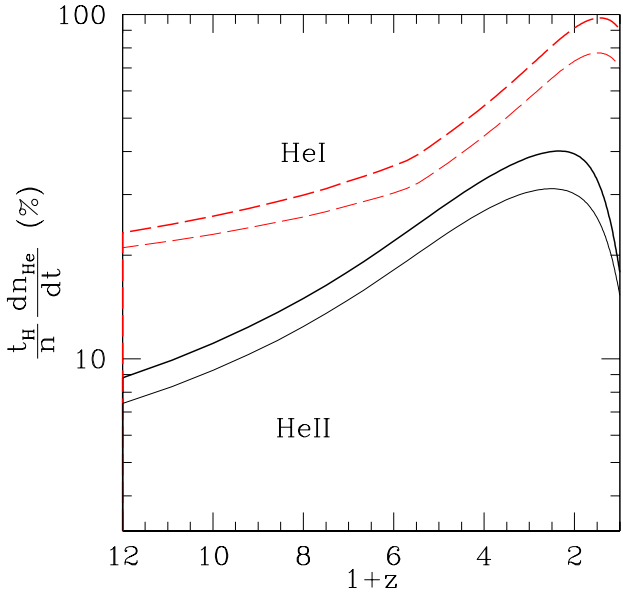,width=8.5cm}}
\caption{\label{fig:He_rateM} Effect of the
  redshifted background on the evolution of the \GI (dashed lines) and
  \GII (solid lines) photoionisation rates, $\Gamma$.  The lines show
  $(t_H/n(He)) d n(He) /dt \sim t_H\Gamma$, where $t_H$ is the Hubble
  time, for model M3 (thick lines) and model M4 (thin lines).}
\end{figure}

In \fig~\ref{fig:helM3} we show the He ionisation history for model
M3. The dotted, solid and dashed lines refer to gas over-densities
$\delta=0.1, 1$, and 10, respectively. Interestingly, in this model
(and in model M4) the \GII was almost fully reionised at high-z and
fully reionised at $z \sim 3$. Helium double ionisation is almost
complete at $z \sim 15$ but afterwards \GII recombines very slowly as
the X-ray emissivity decreases. Note that the recombination rate of
\GIII is four times faster than the one of \HII. The reason for this
slow recombination rate of \GIII is the following.  In this model the
second reionisation of \GII it is not produced by a second peak in the
X-ray emissivity at $z \sim 3$ due to AGN activity.  Instead it can be
explained by the hardening of the spectrum of the background radiation
in the UV bands, that it is produced by the combination of two
effects: (i) the IGM becomes increasingly transparent to ionising
radiation after \HI reionisation at $z_{\rm rei}=7$ and (ii) the
background soft X-rays are redshifted into the hard-UV bands and
ionise more efficiently the \GII.  This effect is a general feature of
all late and intermediate accretion models. This is illustrated in
\fig~\ref{fig:He_rateM}) that shows the effect of the redshifted
background on the evolution of the \GI and \GII photoionisation rates,
$\Gamma$. Neglecting recombinations and the Hubble expansion we have
approximately that $(t_H/n(He)) d n(He) /dt \sim t_H \Gamma$, where
$t_H$ is the Hubble time. The solid lines show $t_H\Gamma(\GII)$ for
\GII and the dashed lines $t_H\Gamma(\GI)$ for \GI as a function of
redshift. The thick lines refer to model M3 and the thin lines to
model M4.

Observations of the \GII \lya forest
\citep[\eg,][]{Reimers:97,Theuns:02} and the line widths of the \HI
\lya forest \citep{RicottiGS:00} at $z \sim 3$ suggest that \GII
reionisation happens at redshift $z \simeq 3$. This is usually
attributed to the observed peak of AGN activity at that redshift, but
as noted in \cite{Miniati:03}, heating from shocks arising from cosmic
structure formation also can make a significant contribution.  The
scenario presented here is another available mechanism that can
explain the observed \GII reionisation at $z \sim 3$ and is consistent
with the redshift evolution of IGM equation of state. Note that in our
models the temperature decreases slowly, monotonically, but does not
show any substantial increase corresponding to the redshift of \GII
ionisation at $z \sim 3$. The same smooth behaviour is exhibited by
the parameter $(\gamma-1)$.


The evolution of the SFR for models M3, M4 and M5 was shown in
\fig~\ref{fig:sfr}. Note that in model M5, in which BH accretion
starts earlier, the thermal feedback strongly suppresses star formation in the
smaller mass galaxies. This effect reduces the number of seed BHs and
therefore the emission of X-rays at high redshift. The filling factor
of \HII regions (shown in the top panels of \fig~\ref{fig:mod1}) has
an evolution similar to that of the SFR: after the initial growth, the
filling factor remains constant at a few percent of the volume. It
starts growing again when most massive galaxies form at $z \sim 15$.
This effect is also observed in the cosmological simulations with
radiative feedback discussed in paper~IIb.

\begin{figure}
\centerline{\psfig{figure=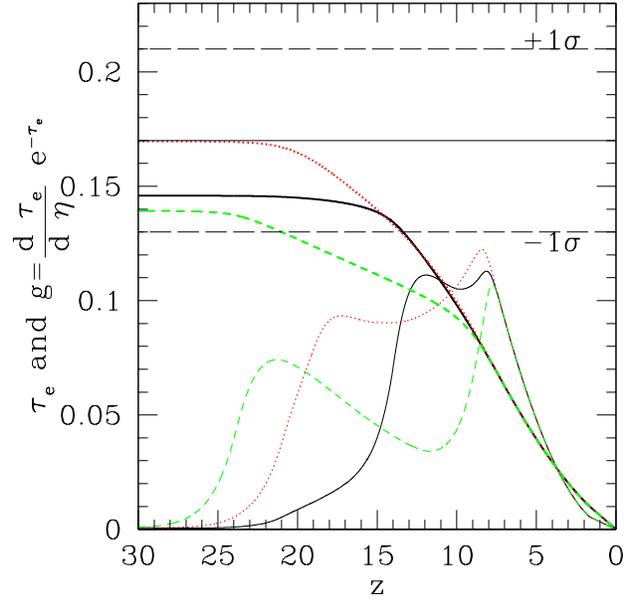,width=8.5cm}}
\caption{\label{fig:mod3} Thomson scattering optical depth, \taue,
  and visibility function, $g(z)$, as a function of redshift. Models
  M3 (solid lines), M4 (dotted lines) and M5 (dashed lines) are shown.
  The models differ for the accretion history of BH seeds (duty cycle)
  but have the same total integrated X-ray energy input. The maximum
  of \taue is obtained for model M4, that has the bulk of accretion at
  intermediate redshift ($z \sim 20$) with respect to models M3 and
  M5.}
\end{figure}

The optical depths to Thomson scattering produced by the X-ray
preionisation models are within the 65\% confidence limit of WMAP. The
models are also in agreement with \taue$\sim 0.103^{+0.060}_{-0.047}$
found by \cite{Tegmark:04} by combining WMAP and other CMB experiments
with the Sloan Digital Sky Survey (see their Table~4).  The optical
depth, \taue, and the visibility function $g(z)=e^{-\tau_{\rm e}}d
\tau_{\rm e}/d \eta$ (where $\eta$ is the conformal time) as a
function of redshift for models M3, M4, and M5 are shown in
\fig~\ref{fig:mod3}.  The three models differ for the accretion
history onto seed black holes. In model M5 most of mass in black holes
is accreted at $z>25$, in model M4 at $z>20$ and in model M3 at
$z>15$. It is interesting that model M4 has the larger \taue, even if
all the models have the same integrated energy input from accreting
BHs. This is due to the small, but non-negligible, recombination rate
of partially ionised gas.  If the IGM is partially ionised at high
redshift and recombinations can be neglected, \taue should increase
the earlier the partial ionisation begins, and vice-versa if the
partial ionisation starts later. But if recombinations are not
negligible at high redshift, then there is an intermediate redshift
that maximises \taue.  In our models this redshift is $z \sim 15$
(model M4).

Another interesting effect worth noticing is the delay between the
formation of the first sources and the build-up of a substantial
background. Roughly the time scale for building the background is
given by a fraction of the Hubble time. For instance, if the first
sources formed at $z=40(30)$, the X-ray background will build up at
$z=25(20)$. Also, as already noted, the age of the universe at $z=30$
is about one Eddington time $t_{\rm Ed}=10^8$ yr. If seed BHs accrete
at the Eddington limit their mass would grow exponentially one
e-folding from $z=30$ to $z=20$ and 4 e-foldings to $z=12$. Therefore,
the energy available to produce X-rays in the very early universe is
limited to 5-10 times the mass of the seed BHs. This means that we
need an increasingly large mass in seed BHs to preionise the IGM
starting at higher redshift (\cf, \fig~\ref{fig:fBHM}). This is only
possible if the IMF of the first stars is top-heavy and has been
discussed in paper~I.

\begin{figure}
\centerline{\psfig{figure=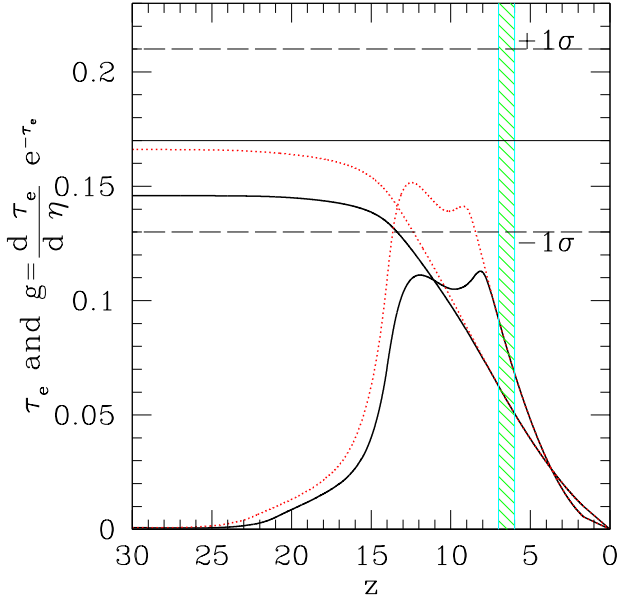,width=8.5cm}}
\caption{\label{fig:mod4} As in \fig~(\ref{fig:mod3}) for model M3
  with spectral index of primordial perturbations $n_{\rm s}=1$ (solid
  lines) and the same model but using $n_{\rm s}=1.04$ (model M3b,
  dot-dashed lines).}
\end{figure}

To illustrate the effect of cosmological parameters on the \taue
produced by X-ray preionisation, in \fig~\ref{fig:mod4} we show \taue
and $g(z)$ for model M3b that has a spectral index of initial density
perturbations $n_{\rm s}=1.04$ and, for comparison, model M3 that has
$n_{\rm s}=1$ normalised to the same amplitude of the power spectrum
at $k=0.05$ h Mpc$^{-1}$ measured by WMAP \citep{Verde:03}. The other
parameters in model M3b and M3 are the same. In model M3b seed BHs
form earlier and the preionisation starts at higher redshift than in
model M3. In this model we get \taue$=0.16$, compared to \taue$=0.14$
of model M3.  This result is qualitatively different if reionisation
is instead produced by UV sources. In this second case, in paper~I we
have shown that \taue is not substantially increased if the power
spectrum of initial perturbations has more power on small scales
because star formation in small mass galaxies is regulated and limited
by feedback effects.


\begin{table}
\caption{Results of the semianalytic models}\label{tab:2}
\begin{tabular*}{8.3 cm}[]{l|cccccc}
Model & \taue & $z_g^{max}$ & $y$-par &
${\omega_{\rm BH} \over \omega_*}$ &  $f_{\rm XRB}$ & $\Gamma(\GII) t_{\rm
  H}$\\
 &  &  & $\times 10^{7}$ & (\%)& (\%) & (\%)\\
\hline
M3 & 0.146 & 12.0 & $3.64$ & 3  & 5 (12) & 35 \\
M3b & 0.166 & 12.5 & $4.26$ & 3  & 5 (12) & 35 \\
M4 & 0.170 & 17.0 & $3.95$ & 9  & 7 (17) & 25 \\
M4b & 0.170 & 17.0 & $3.95$ & 9  & 7 (17) & 25 \\
M5 & 0.140 & 21.5 & $2.96$ & 42 & 0.3 (1)& 1.5\\
\end{tabular*}
\\

Meaning of the values in each column: \taue is the Thomson scattering
optical depth; $z_g^{max}$ is the
high-redshift maximum of the visibility function $g(z)$; $y$ is the
Compton distortion parameter; $\omega_{\rm BH}/\omega_*$ is the maximum ratio of
BH to stellar cosmic density (to translate $\omega_{\rm BH}$ to
$\rho_{\rm BH}$ multiply by $5.5 \times 10^9$ M$_\odot$ Mpc$^-3$); 
$f_{\rm XRB}$ is the fraction of the X-ray background at 50-100
keV (and 2-10 keV, in parenthesis) due to early black holes; $\Gamma(\GII) t_{\rm
  H} \sim (t_H/n_{\GII}) dn_{\GII} /dt$ is the fractional rate of
  \GII photoionisation per Hubble time per helium atom due to the
  redshifted X-ray background.
\end{table}

In conclusion, adopting the model constraints listed at the beginning
of this section, we obtain values of \taue consistent with WMAP. The
models are consistent with and do not violate any observation at $z<6$
that are evident. In the following paragraphs we discuss a few obvious
observational constraints. But we will focus and return to this
discussion in paper~IIb.

Reheating of the IGM at redshifts $z<10^5$, produces a Comptonised spectrum of
the CMB blackbody. The Compton distortion is usually parametrised by
the value of the $y$:
\[
y \equiv \int_0^{\tau_e^{\rm max}} {k_{\rm B}(T_e -T_{\rm CMB}) \over m_e c^2}d
\tau_e \simeq 1.7 \times 10^{-6} \int_0^{\tau_e^{\rm max}} {T_e \over
  10^4 K}d \tau_e,
\]
where $T_e$ is the electron temperature, $T_{\rm CMB}$ is the CMB photon
temperature and $\tau_e$ is the optical depth to electron Compton
scattering.  An upper limit on the $y$-parameter $y \simlt 1.5 \times
10^{-5}$ at 95\% CL has been determined using COBE satellite data
\citep{Fixsen:96}. The maximum value of $y$ that we find in our
models, $y \simeq 4 \times 10^{-7}$, is not constrained by the current
data on the spectral distortion. A future experiment with an upper
limit on the $y$-parameter improved by two orders of magnitude would
be able to detect a possible X-ray preionisation and reheating in the
early universe.
\begin{figure}
\centerline{\psfig{figure=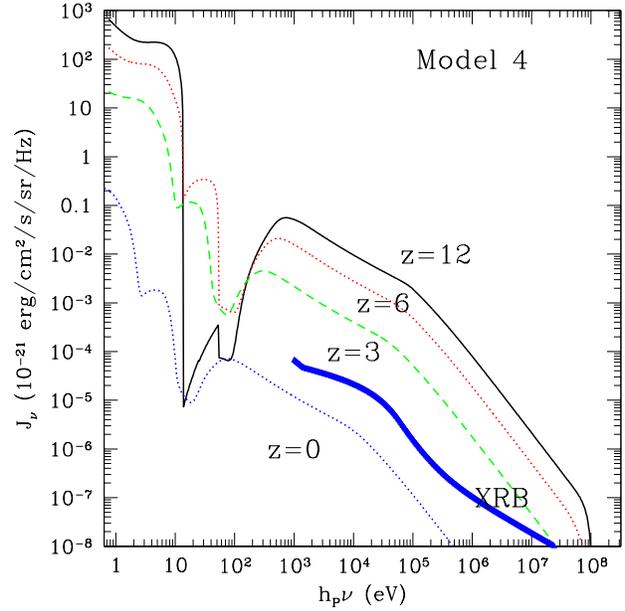,width=8.5cm}}
\caption{\label{fig:J21M4} Background radiation at $z=12, 6, 3$ and
  $z=0$ for model M4. The thick line shows the X-ray background
  radiation at $z=0$ \protect{\citep{Comastri:95}}. The redshifted
  X-ray background produced by early epoch of black hole accretion
  constitutes about $5-10$\% of the hard X-ray background in the 2-50
  keV bands.}
\end{figure}

In \fig~\ref{fig:J21M4} we show the background radiation at $z=12, 6,
3$ and $z=0$ for model M4. The thick line shows the X-ray background
radiation at $z=0$ \protect{\citep{Comastri:95}}. We see that the
redshifted X-ray background produced by early epoch of black hole
accretion constitutes about $5-10$\% of the hard X-ray background in
the 2-50 keV bands.

The observed (X,$\gamma$) ray background is produced almost completely
by known sources (mainly Seyfert galaxies at $z \sim 1-2$).  However
there is some room for additional sources and recently
\cite{deLuca:03} have estimated that that about 20\% of the observed
background may be produced by a new population of faint X-ray sources,
currently undetected within the sensitivity limits of the deepest
X-ray surveys.  The intensity of the high redshift X-ray background
decreases from its value at redshift $z$ as $(1+z)^{-3+\phi}$, where
$\phi$ is the logarithmic slope of the spectral energy distribution.
By imposing that the redshifted background produced at the redshift of
the peak of the BH accretion does not exceed 20\% of the observed
value at $z=0$ we find the following upper limit for the BH accretion
rate:
\begin{equation}
\dot \rho_{\rm ac}^{\rm max} \simlt (10^{-2} M_\odot {\rm Mpc}^{-3}
{\rm yr}^{-1})\left({0.1 \over \beta}\right)\left({1+z \over 20}\right)^{-3.7}.
\label{eq:maxac}
\end{equation}
Note that we this expression has been derived for the accretion
history of the models presented in the present paper. It implicitly
assumes that $\dot \rho_{\rm ac}=\Delta\ rho_{\rm ac}/ \Delta t$ with
$\Delta t \simeq t_H(z)$, where $t_H(z)$ is the Hubble time at the
redshift of maximum accretion (\ie, the accretion rate rapidly fades
after the maximum).

In \tab~\ref{tab:2} we summarise the values of some relevant
quantities computed from the semianalytic models listed in
\tab~\ref{tab:1}.  See the table footnote for the explanation of the
meaning of the quantities shown in each column of the table.


\section{Discussion and Summary}\label{sec:conc}

In a previous paper (paper~I), we showed that even with a top heavy
mass function \pop3 stars could not reionise the universe to a
significant degree if they end their lives as pair instability
supernovae. The reason is that such events contaminate the high
density regions to such an extent that further formation of extremely
low metallicity stars becomes impossible before significant ionisation
has occurred.  But, if these same stars implode to massive black
holes, then accretion onto them at near Eddington rates will produce a
pervasive X-ray background which can quite efficiently partially
preionise the bulk of the cosmic baryons before $z=15$ and give an
optical depth to Thompson scattering (\cf, \tab~\ref{tab:2})
consistent with WMAP results.  The primary reason for the increased
efficiency is the long mean free path of X-rays which both increases
the fraction of photons that escape the parent halo and ionises
preferentially in the low density regions where recombination rates
are low. In addition X-rays only partially ionise the IGM to 10-20\%,
therefore the recombination rate is reduced with respect to the
reionisation scenario produced by UV sources. Thus we find that the
average number of times a baryon is photoionised in this picture is
about unity for \taue$\simeq 0.15$ while it is about ten in the case
when stellar UV is used (paper~I). Moreover, since the X-rays
preionise the low density regions of the IGM at early times, the
required number of ionising photons from \popII stars to have
reionisation at redshift $z \simeq 6$ is much reduced.

But we might ask, could the X-rays arise as easily from X-ray binaries
or could SN remnants \citep[\cf,][]{Oh:00}?  These other sources are
certainly helpful but they will generate somewhat less energy in
X-rays than accretion can do. The ratio of X-ray energy to massive
star formation expressed as an efficiency is $\epsilon_{\rm SN}=E_{51}
10^{51} {\rm erg}/143 M_\odot c^2 \sim 4 \times 10^{-6}$ for these
sources (assuming \popII with Salpeter IMF and neglecting Compton
cooling of the X-ray emitting hot gas) but we see from $\epsilon_{\rm
  ac}=0.15 \beta (M_{\rm SMBH}/M_{\rm bulge})$ that the comparable
number from accretion is $2.25 \times 10^{-5}$.

What are the observable consequences of the scenario we are proposing
(in addition to generating a large \taue)?
\begin{enumerate}
\item The power spectrum of the cross correlation of temperature and
  polarisation of the CMB is sensitive to the visibility function
  $g(z)=e^{-\tau_{\rm e}}d \tau_{\rm e}/d \eta$ (\cf,
  \fig~\ref{fig:mod3}-\ref{fig:mod4}). The Planck satellite should be
  able to distinguish between the visibility function produced by an
  early X-ray partial ionisation (peaking at large z) or the one
  expected for reionisation by stellar sources (peaking at lower z).
\item Since the assumed spectrum (\fig~\ref{fig:sed}) has a positive
  slope $(d\ln L_\nu /d\ln \nu) >3$ for much of the relevant range of
  frequencies there are ever fresh ionising photons available and the
  probability per unit time that a neutral hydrogen atom is ionised
  does not decrease with time. This produce softer photons that, even
  if the ionisation efficiency of X-rays decreases when the ionisation
  fraction exceeds 10-20\%, enhance the ionisation fraction in the
  voids to values larger than 20 \% because of the contribution from
  the redshifted photons of the X-ray background.  In most models, the
  photons from the redshifted X-ray background can fully reionise \GII
  at redshift $z \sim 3$ without any additional contribution from
  quasars at lower redshifts. Due to the heating rate by \GII
  ionisation the temperature of the mean density intergalactic medium
  remains close to $10^4$ K down to redshift $z \sim 1$ and the
  IGM equation of state has index $(\gamma-1) \sim 0.3$.
\item The maximum accretion rate onto seed black holes as a function
  of redshift is limited by the observed magnitude of the $\gamma$ ray
  background at $z=0$. This limits the ability of late preionisation
  models to produce large values of \taue and contribute substantially
  to the mass growth via accretion of seed black holes. In models that
  produce optical depth \taue$\sim 0.17$ consistent with WMAP, the
  redshifted X-ray background produced by the early epoch of black
  hole accretion constitutes about $5-10$\% of the hard X-ray
  background in the 2-50 keV bands.
\item The model predicts that dwarf spheroidal galaxies, if they are
  preserved fossils of the first galaxies, would host a mass in black
  holes that is 5-40\% of their stellar mass (\cf, \tab~\ref{tab:2}).
  The larger values are produced in the early preionisation scenarios
  and it might be possible to rule them out on the base of
  observations from dynamical considerations.
\item The IGM is reheated to $T =10^4$ K before the ionisation
  fraction exceeds 10\% producing typical values of the Compton
  distortion parameter $y \sim 4 \times 10^{-7}$.  The
  observational upper limit on this parameter is $y < 1.5 \times
  10^{-5}$, too large to constrain our models. Star formation in
  the smaller mass haloes is reduced by the increase in the IGM Jeans
  mass following this early reheating.  
\item The redshifted 21cm signal in emission and absorption against
  the CMB can be used to discriminate between preionisation scenarios
  and reionisation from UV sources. This will be discussed in
  paper~IIb, where we will also show that preionisation by X-rays
  produces CMB secondary anisotropies on small angular scales
  that are easy to recognise when compared to models of inhomogeneous
  early reionisation by stellar sources with the same \taue.
\end{enumerate}

What are the requirements of the X-ray preionisation models in terms
of seed black hole production from the first stars and accretion? 

Since the time scale for accretion, $t_{\rm Ed}=10^8$ yr, is
comparable to the Hubble time at $z=30$ the maximum X-ray emissivity
at high-redshift is proportional to the total mass of the seed BHs.
For this reason, in order to get a larger \taue in early preionisation
scenarios, a top-heavy IMF is favoured over a Salpeter IMF. But this
is true only if super-massive stars mostly end their lives collapsing
directly into BHs without exploding as SNe or as pair-instability SNe.

We have shown models where the accretion onto seed black holes is near
the Eddington rate at early redshift. This might perhaps be difficult
to achieve.  But it is worth noticing that it is not strictly
necessary that the accretion is at the Eddington rate initially.
Although, if the accretion rate is less efficient than that, a larger
stellar mass fraction producing seed black holes, $f_{\rm BH}$, is
needed to accommodate for this (see \eq~(\ref{eq:ombh})).

In this paper we have explored a range of models with different
accretion histories onto seed black holes.  But we note that the
models are not all equally plausible scenarios.  The very early
preionisation scenario (M5) is not the most favourable for several
reasons: (i) as noted by \cite{MadauR:03}, accretion in small-mass
haloes can be difficult because of the substantial gas
photoevaporation in these galaxies (ii) if seed BHs do not accrete at
near the Eddington rate the mass fraction, $f_{\rm BH}$, of seed BHs
is extreme also assuming top-heavy \pop3 stars (iii) the number of
recombinations per ionising photon in this model is larger than in the
intermediate preionisation model and therefore \taue is reduced.  The
late preionisation model (M3) is also not favourable because it
produces a low \taue if we constrain the maximum black hole accretion
rate using \eq~(\ref{eq:maxac}) to be consistent with the observed
$\gamma$-ray background at $z=0$.  We therefore conclude that
preionisation starting at $z \sim 15-20$ (\eg, model M4) is the most
plausible scenario to reproduce \taue measured by WMAP. In this case
$\sim 10$\% of the X-ray background in the 2-50 keV band is produced
by the early generation of mini-quasars and roughtly half of the
currently estimated black hole mass density was formed at early times
(even if only a fraction of them may have merged into the SMBHs in the
galactic centres).  In paper~IIb we study this scenario using
hydrodynamic cosmological simulations with radiative transfer and we
discuss observations that could probe this model.

We also note in conclusion that this picture while consistent with the
central values of WMAP results could not easily produce \taue at the
upper end of the presently allowed values. If \taue$ \simgt 0.2$, X-ray
preionisation is not likely the dominant responsible mechanism and
other scenarios need to be considered (\eg, decaying particles, non
Gaussian perturbations, spectral index $n_{\rm s}>1$, primordial BHs,
unconventional recombination).

\subsection*{ACKNOWLEDGEMENTS}

MR is supported by a PPARC theory grant. Research conducted in
cooperation with Silicon Graphics/Cray Research utilising the Origin
3800 supercomputer (COSMOS) at DAMTP, Cambridge.  COSMOS is a UK-CCC
facility which is supported by HEFCE and PPARC. MR thanks Martin
Haehnelt and the European Community Research and Training Network
``The Physics of the Intergalactic Medium'' for support.  The authors
would like to thank Andrea Ferrara, Nick Gnedin, Martin Haehnelt,
Piero Madau and Martin Rees for stimulating discussions and Mark
Dijkstra for noticing some incorrect numbers in the first draft of
this paper. MR thanks Erika Yoshino support.

\appendix

\section{A Semianalytic Model for Reionisation}\label{ap:A}

We implemented a semianalytic model to study reionisation, chemical
evolution and re-heating of the IGM. The code is based on the method
of \cite{Chiu:00} for the calculation of the filling factor of ionised
regions and for the star formation recipe. The main difference is the
inclusion of radiative transfer for the background radiation that
allows us to follow the thermal and chemical evolution of the IGM
outside the \HII regions surrounding each UV source. We also consider
the thermal feedback produced by the X-ray reheating of the IGM (see
\S~\ref{ap:feed}).  The mass function of DM haloes and their
formation/merger rates are calculated using the extended
Press-Schechter formalism.  The star formation rate is assumed to be
proportional to the formation rate of haloes (see \S~\ref{ap:sf}).  We
consider the IGM as a two-phase medium: one phase is the ionised gas
inside the \HII regions and the other is the neutral or partially
ionised gas outside.  We solve the radiative transfer for the volume
averaged specific intensity and we derive the specific intensity
inside and outside the \HII regions from their volume filling factor
and by separating the contribution from local UV sources and the
background radiation from distant sources (see \S~\ref{ap:radhII}).
We solve the time-dependent chemical network for eight ions and the
thermal evolution as a function of the gas density outside the \HII
regions. Given the clumping factor of fully ionised gas around the UV
sources we evolve the filling factor, temperature and chemistry inside
the \HII regions (see \S~\ref{ap:chem}).  The following heating and
cooling processes are included: collisional- and photo-ionisation
heating, ionisation by secondary electrons (see \S~\ref{ap:secel}), H,
He, \H2 cooling, Compton and adiabatic cosmological cooling. The rates
are the same as the ones used in \citep{RicottiGS:01}.  For some test
cases the results of the semianalytic code are in good agreement with
the results of the cosmological simulations with radiative transfer
presented in this paper and in \cite{RicottiGSb:02}.
 
\subsection{Star Formation}\label{ap:sf}

We assume that stars and quasars form in virialised DM haloes. We use
the extended Press-Schechter formalism to calculate the formation rate
of bound DM haloes of a given mass $M_{\rm dm}$ at the time $t$. In
general, given a probability distribution function (PDF) of density
perturbations, $P(\nu)$, the DM comoving mass fraction of collapsed
objects with mass between $M_{\rm dm}$ and $M_{\rm dm}+dM_{\rm dm}$ is given by
$\Omega_{\rm ps}d\ln M_{\rm dm}=P(\nu)d\nu$, where
\[
\int_0^\infty d\ln M_{\rm dm} \Omega_{\rm ps}=\int_0^\infty d\nu P(\nu)=1.
\]
The comoving number density of collapsed objects
$N_{\rm ps}=\Omega_{\rm ps}\rho_0/M_{\rm dm}$, where $\rho_0$ is the mean DM
density at $z=0$, is given by
\begin{eqnarray}
N_{\rm ps} dM_{\rm dm} &=& 2 \rho_0 P(\nu) \nu \Delta(M_{\rm dm}) d\ln
M_{\rm dm},\\
\nu &=& {\delta_c \over \sigma(M_{\rm dm}) D(t)},\\
\Delta(M_{\rm dm}) &=& {d \ln \nu \over d \ln M_{\rm dm}}=-{d \ln
  \sigma(M_{\rm dm}) \over d \ln M_{\rm dm}},
\end{eqnarray}
where $\sigma(M_{\rm dm})$ is the variance of the fluctuation on a scale
$M_{\rm dm}$ linearly extrapolated to $z=0$, $\delta_c \approx 1.69$ is
the linear overdensity for collapse of a top-hat perturbation and
$D(t)$ is the linear growth factor. Here we use a Gaussian PDF $P
=(2\pi)^{-1/2} \exp{(-\nu^2/2)}$, but the method has been applied by
\citep{Chiu:00} to non-Gaussian PDFs. With a similar procedure as in
\citep{Ricotti:02}, we normalise the star formation efficiency by
assuming that at $z=0$ a fraction of baryons $\omega_*=14$\% has been
converted into stars. The global SFR between redshift $0<z<5$ agrees
within the errors with the observed global SFR
\citep[\eg,][]{Lanzetta:02}. As shown in \S~\ref{sec:xsour}, a similar
method of normalisation is applied to the BH accretion efficiency by
assuming that at $z=0$ a fraction of baryons $\omega_{\rm BH}=10^{-4}$ is
in black holes.

The formation rate of virialised objects, $\dot N_{\rm form}$, can be
derived using the Press-Schechter formalism and calculating the rate
of ``destruction'' of bound objects that are incorporated in larger
haloes:
\begin{equation}
\dot N_{\rm form} = \dot N_{\rm ps} + \dot N_{\rm dest}.
\label{eq:ap0}
\end{equation}
As in \cite{Chiu:00}, we use a destruction probability $\phi(t)=
\dot N_{\rm dest}/N_{\rm ps}=\dot D/D$ derived assuming that the destruction
probability is scale-invariant \citep[see ][]{Sasaki:94}. It follows that
the probability that an object formed at time $t_f$ exists at
time $t$ is 
\begin{equation}
p_{\rm surv}(t|t_{\rm f})=\int_t^{t_{\rm f}} dt \phi(t)= {D(t) \over
  D(t_{\rm f})},
\end{equation}
where $D(t)$ is the linear growth factor. At redshifts $z>1$,
$p_{\rm surv}(t|t_{\rm f})=a(t) /a(t_{\rm f})$ for the concordance $\Lambda$CDM
cosmology. From \eq~(\ref{eq:ap0}), recalling that $\dot D/D \approx
\Omega^{0.6} (\dot a /a)$, it follows
\[
\dot N_{\rm form}= -N_{\rm ps}{\nu \over P}{d P  \over d\nu}
\phi(t)=N_{\rm ps}\nu^2 \Omega^{0.6}\left({\dot a \over a}\right),
\]
where $d\ln P/d\nu=-\nu$ for a Gaussian PDF ($P\propto
\exp{(-\nu^2/2)}$).  The comoving number density of haloes at time $t$
that formed at time $t_f$ is given by
\[
\dot N(M_{\rm dm}, t_f, t)dM_{\rm dm} dt_{\rm f}=\dot N_{\rm
  form}(M_{\rm dm}, t_f)p_{\rm surv}(t|t_{\rm f})dM_{\rm dm} dt_{\rm f}.
\]
The emission rate per unit volume from the sources, often called the
source function $\overline S(t)$, when expressed in dimensionless
units ${\cal S}=\overline S(t)/(n_H h\nu_0)$, where $h\nu_0=13.6$ eV
and $n_H$ is the hydrogen mean number density, is given by
\begin{equation}
\begin{split}
{\cal S}(t)=n_0^{-1}\int_0^{\infty} dM_{\rm dm} \int_0^t dt_{\rm f}\\
\times \Psi(M_{\rm dm},t_{\rm f}) \dot N((M_{\rm dm}, t_f,t){\cal
  L}(M_{\rm dm}, t_{\rm f}, t).
\end{split}
\label{eq:ap1}
\end{equation}
The function $\Phi(M_{\rm dm},t_{\rm f})$, defined in \S~\ref{ap:feed}, is a
step function that determines the minimum mass of forming galaxies as
a function of the their time of formation, $t_f$. The dimensionless
luminosity from each source is ${\cal L}_\nu \propto g_\nu
\epsilon_{\rm UV} M_* c^2 x \exp{(-x}/t_{\rm dyn})$, where $g_\nu$ is the SED
of the source, $x=(t-t_{\rm f})/t_{\rm dyn}$ and $t_{\rm dyn}=(3\pi /32
G \rho_{\rm vir})$ is the free-fall dynamical time.

\subsection{Cooling and Chemistry}\label{ap:chem}

We solve the time-dependent equations for the photo-chemical
formation/destruction of eight chemical species (H, H$^+$, H$^-$, H$_2$,
H$_2^+$, He, He$^+$, He$^{++}$), including the 37 main processes
relevant to determine their abundances \citep{ShapiroKang:87}.  We use
ionisation cross sections from \cite{Hui:97} and photo-dissociation
cross sections from \cite{Abel:97}.  We solve the energy conservation equation 
\begin{equation}
{d E_{\rm gas} \over dt}=\Gamma-\Lambda ,
\end{equation}
where $E_{\rm gas}=(3kT/2) n_H(1+x(He)+x_e)$, with $n_H$ the hydrogen
number density, and $x(He)$ and $x_e$ the helium and electron
fractions, respectively. Note that $n_H$ is a function of time because
of Hubble expansion. Also, $x_e$ is time-dependent; neglecting this
effect results in a temperature that is overestimated by about a
factor of two.  The cooling function, $\Lambda$, includes H and He
line and continuum cooling \citep{ShapiroKang:87}, \H2 rotational and
vibrational cooling excited by collisions with H and \H2
\citep{Martin:96,Galli:98} and adiabatic cosmic expansion cooling. The
heating term, $\Gamma$, includes Compton heating/cooling and
photoionisation/dissociation heating.  We solve the system of ODEs for
the abundances and energy equations, using a $4^{\rm th}$-order
Runge-Kutta solver. We switch to a semi-implicit solver
\citep{GnedinG:98} when it is more efficient (\ie, when the abundances
in the grid are close to their equilibrium values).  The spectral
range of the radiation is between 0.7 eV and 10 keV. The primordial
helium mass fraction is $Y_P=\rho(He)/\rho_b=0.24$, where $\rho_b$ is
the baryon density, so that $x(He)=Y_P/4(1-Y_P)=0.0789$. The initial
values at $z=z_i$ for the temperature and species abundances in the
IGM are: $T=10$ K, $x_{\rm H_2}=2 \times 10^{-6}$, and $x_e \simeq
x_{H^+}=10^{-5}/(h\Omega_b\Omega_0^{1/2})=6.73 \times 10^{-4}$.  The
initial abundance of the other ions is set to zero.

Outside the Str\"omgren spheres surrounding the UV sources the thermal
and chemical histories are calculated as a function of the gas
density, using the specific intensity $\langle J_\nu \rangle_{\rm
  back}$ of the background radiation defined in \S~\ref{ap:back}.  In
this paper we show the abundances and the temperature of the IGM for
three values of the baryon overdensity $\delta=\rho / \rho_0=0.1, 1,
10$.  Inside the \HII regions we calculate the abundances of only H
and He ions, since the abundances of molecular hydrogen and its ions
are negligible, being photodissociated by \HI ionising radiation. The
effective coefficients of collisional ionisation and recombinations
are larger by a factor $C_{\HII}$, to take into account the gas
clumping inside the \HII regions (see \S~\ref{ap:radhII}).  The
photoionisation and photoheating rates are calculated as shown in
\S~\ref{ap:secel} using the specific intensity $J_{\HII}$ of the
radiation field inside \HII regions.  The clumping factor $C_{\HII}$
is tabulated as a function of redshift by fitting the clumping in the
cosmological simulations with radiative transfer presented in paper~IIb.

\subsubsection{Secondary Ionisation and Heating from X-rays}\label{ap:secel}

Photoionisation of \HI, \GI, and \GII by X-rays and EUV photons
produces energetic photoelectrons that can excite and ionise atoms
before their energy is thermalised. This effect can be important
before reionisation \citep{Oh:00, Venkatesan:01}, when the gas is
almost neutral and the spectrum of the background radiation is hard
due to the large optical depth of the IGM to UV photons.

Collisional ionisation and excitation of \GII by primary electrons are
neglected since, in a predominantly neutral medium, they are
unimportant.  The primary ionisation rate for the species $i=\HI,
\GI,\GII$ is,
\begin{equation}
\zeta^i= 4\pi \int_{\nu_i}^\infty {J_\nu \over h_p \nu}\exp(-\tau_\nu)\sigma_\nu^i d\nu,
\end{equation}
where $\tau_\nu$ is the continuum optical depth, and $\sigma_\nu^i$ is
the photoionisation cross section of the species $i$.  Secondary
ionisation enhances the photoionisation rates as follows:
\begin{eqnarray}
\zeta_s^{\HI} = \zeta^{\HI}+\sum_{i=\HI,\GI,\GII} \zeta^i\langle
\Phi^{\HI}(E_0^i,x_e)\rangle \label{eq:z1}\\
\zeta_s^{\GI} = \zeta^{\GI}+\sum_{i=\HI,\GI,\GII} \zeta^i\langle
\Phi^{\GI}(E_0^i,x_e)\rangle,
\end{eqnarray}
where $\langle \Phi^{\HI}(E_0^i,x_e)\rangle$ and $\langle
\Phi^{\GI}(E_0^i,x_e)\rangle$ express the average number of secondary
ionisation per primary electron of energy $E_0^i = h_p \nu - I^i$
weighted by the function $(J_\nu/h_p \nu)
\exp(-\tau_\nu)\sigma_\nu^i$.  Here $I^i=h_p \nu_i$ is the ionisation
potential for the species $i$.

The photoionisation heating rates for the species $i=\HI, \GI$ are given by,
\begin{equation}
\Gamma^i= 4\pi \int_{\nu_i}^\infty {J_\nu \over h_p \nu}\exp(-\tau)\sigma_\nu^i
E_h(E_0^i,x_e) d\nu.\label{eq:heat}
\end{equation}
Analytic fits to the functions $\Phi^{\HI}, \Phi^{\GI}$ and $E_h$,
based on the Monte Carlo results of \cite{ShullVan:85} are given in
Appendix~B in \cite{RicottiGSa:02}.

\subsection{Radiative transfer}\label{ap:radhII}
The evolution of the specific intensity $J_\nu$ [erg cm\mmm s\m Hz\m
sr\m] of ionising or dissociating radiation in the expanding universe,
with no scattering, is given by the following equation:
\begin{equation}
        {\partial J_\nu\over\partial t} +
        {\partial\over\partial x^i}\left(\dot{x}^i J_\nu\right) -
        H\left(\nu{\partial J_\nu\over\partial\nu}-3J_\nu\right) =
        - k_\nu J_\nu + S_\nu.
        \label{eq:Jnueq}
\end{equation}
Here, $x^i$ are the comoving coordinates, $H$ is the Hubble constant,
$k_\nu$ is the absorption coefficient, $S_\nu$ is the source function,
and $\dot{x}^i=cn^i/a$, where $n^i$ is the unit vector in the
direction of photon propagation and $a=(1+z)^{-1}$ is the scale factor.
The volume-averaged mean specific intensity is
\begin{equation}
        \bar J_\nu(t) \equiv \langle J_\nu(t,\vec{x},\vec{n})\rangle_V,
        \label{eq:Jnubardef}
\end{equation}
where the averaging operator acting on a function $f(\vec{x},\vec{n})$
of position and direction is defined as:
\begin{equation}
        \langle f(\vec{x},\vec{n})\rangle_V =
        \lim_{V\rightarrow\infty} {1\over 4\pi V}
        \int_V d^3x \int d\Omega f(\vec{x},\vec{n}).
        \label{averagoper}
\end{equation}
The mean intensity $\bar J_\nu(t)$ satisfies the following equation:
\begin{equation}
        {\partial \bar J_\nu\over\partial t} -
        H\left(\nu{\partial \bar J_\nu\over\partial\nu}-3\bar J_\nu\right) =
        - \bar k_\nu \bar J_\nu + \bar S_\nu,
        \label{eq:back}
\end{equation}
where, by definition, $\bar S_\nu \equiv \langle S_\nu \rangle_V$, and
$\bar k_\nu \equiv \langle k_\nu J_\nu\rangle_V / \bar J_\nu$.  In
general, $\bar k_\nu$ is not a space average of $k_\nu$, since it is
weighted by the local value of the specific intensity $J_\nu$. But
when the mean free path of radiation at frequency $\nu$ is much larger
than the characteristic scale of the problem (in our case the size of
the computational box), we have $\bar k_\nu= \langle k_\nu \rangle_V$.

If we rewrite \eq~(\ref{eq:back}) in terms of the dimensionless
comoving photon number density $n_\nu = (aL_{\rm box})^3 4\pi \bar J_\nu/
h_p$ where $L_{\rm box}$ is a spatial scale (in this paper we take it
to be the comoving size of the computational box), and $h_p$ is the Planck
constant, using substitutions $dt= (a^2/H_0)d\tau$ and $\xi =
\ln(\nu)$, \eq~(\ref{eq:back}) can be reduced to the following
dimensionless equation:
\begin{equation}
  {\partial n_{\xi} \over \partial \tau} = {1 \over a}{d a \over d \tau} {\partial
  n_{\xi} \over \partial \xi} - \alpha_{\xi}
  n_{\xi} + S_{\xi},
\label{eq:back2}
\end{equation}
where $\alpha_\xi = a^2 \bar k_\nu/H_0$ and $S_\xi = a^5L_{\rm box}^3
4\pi \bar S_\nu/h_p H_0$.  Using a comoving logarithmic frequency variable,
\begin{equation}
  \bar \xi = \xi + \ln(a),
\end{equation}
equation~(\ref{eq:back2}) can be reduced to
\begin{equation}
  {\partial n_{\bar \xi} \over \partial \tau} = - \alpha_{\bar \xi}
  n_{\bar \xi} + S_{\bar \xi},
\end{equation}
which has the formal solution,
\begin{equation}
\begin{split}
  n_{\bar \xi}(\tau+\Delta \tau) = n_{\bar \xi}(\tau)\exp\left[- 
    \int_\tau^{\tau+\Delta
      \tau} dt^\prime \alpha_{\bar
      \xi}(t^\prime)\right]\\
+ \int_\tau^{\tau+\Delta \tau}dt^\prime
  S_{\bar \xi}(t^\prime)\exp\left[-\int_{t^\prime}^{\tau+\Delta \tau} 
    d\tau^\prime \alpha_{\bar
      \xi}(\tau^\prime)\right].
\end{split}
\label{eq:formal}
\end{equation}
We calculate \eq~(\ref{eq:formal}) at each time step, $\Delta
\tau$, of the simulation.  The two integrals inside the square
brackets on the right side of \eq~(\ref{eq:formal}) can be solved
analytically. We solve the third integral numerically. 

\subsubsection{Background radiation and radiation inside \HII regions}\label{ap:back}

Two terms contribute to the volume averaged specific intensity in
\eq~(\ref{eq:formal}).  The first term is produced by the background
radiation from redshifted distant sources and the second term is
produced by the local sources:
\begin{equation}
\bar J_\nu = \bar J_\nu^{\rm sour} + \bar J_\nu^{\rm back}.
\label{eq:apb1}
\end{equation}
If the emissivity of the sources is zero (${\cal S}=0$), it follows that
$\bar J_\nu^{\rm sour}=0$. Only photons with a mean free path
larger than the mean distance between the sources contribute to the
background radiation, $\bar J_\nu^{\rm back}$. 

If the volume filling factor of the Str\"omgren spheres surrounding
the UV source is $f_{\HII}$, we can rewrite the volume averaged
specific intensity as the sum of the mean specific intensity inside
the volume occupied by the \HII regions and the volume outside the
\HII regions:
\begin{equation}
\bar J_\nu= \bar J_\nu^{\HII} f_{\HII} + \bar J_\nu^{\rm back} (1-f_{\HII}).
\label{eq:apb2}
\end{equation}
From \eq~(\ref{eq:apb1}) and  \eq~(\ref{eq:apb2}) we get,
\begin{equation}
\bar J_\nu^{\HII}= {\bar J_\nu^{\rm sour} \over f_{\HII}} + \bar J_\nu^{\rm back}. 
\end{equation}
In order to derive the specific intensity, $\bar J_\nu^{\HII}$, inside
the Str\"omgren spheres we need to estimate their volume filling
factor $f_{\HII}$. We calculate the evolution of the filling factor of
\HII regions following closely the method introduced by
\cite{Chiu:00}.  For the sake of completeness we show the equations
that we solve to derive $f_{\HII}$ in the next section.

\subsubsection{Filling factor of \HII regions}

In order to calculate $f_{\HII}$ it is more convenient to introduce
the porosity parameter, $Q(t)$, of the \HII regions defined by the
relationship $f_{\HII}=1-\exp{[-Q(t)]}$. The porosity $Q(t)$ offers
the advantage that when $f_{\HII} \ll 1$ we have $Q(t) \approx
f_{\HII}$ and when $f_{\HII} \simeq 1$ we have $Q(t) \simgt 1$. The
porosity of the \HII regions is given by
\begin{equation}
\begin{split}
Q(t)=\int_0^{\infty} dM_{\rm dm} \int_0^t dt_{\rm f}\\
\times \Psi(M_{\rm dm},t_{\rm f}) \dot N((M_{\rm dm},
t_f,t)V_{\HII}(M_{\rm dm}, t_{\rm f}, t),
\end{split}
\label{eq:ap2}
\end{equation}
where $V_{\HII}=(4 \pi/3) r_{\HII}^3$ is the comoving volume filled by the
Str\"omgren spheres, where $r_{\HII}$ is their comoving radius.
Comparing \eq~(\ref{eq:ap1}) to \eq~(\ref{eq:ap2}) we find the
following relationship between the ionised volume per unit luminosity
and the mean comoving luminosity density ${\cal S}$:
\begin{equation}
{n_0 V_{\HII} \over {\cal L}(t)}={Q(t) \over {\cal S}(t)}.
\label{eq:ap3}
\end{equation}
The contribution of each source to the mean specific energy density,
$E=4 \pi \overline J_\nu /c$ is given by, 
\begin{equation}
\begin{split}
\overline E={1 \over c} \int_0^{\infty} dM_{\rm dm} \int_0^t dt_{\rm f}
\Psi(M_{\rm dm},t_{\rm f}) \dot N(M_{\rm dm}, t_f, t) \int {dV L \over 4 \pi (a r)^2}\\
={1 \over c} \int_0^{\infty} dM_{\rm dm} \int_0^t dt_{\rm f} \dot
N(M_{\rm dm}, t_{\rm f}, t) L r_{\HII} a(t)^{-2}.
\end{split}
\end{equation}
Using \eq~(\ref{eq:ap3}) we find that, in dimensionless units
${\cal E}=\overline E/(n_H h\nu_0)$, the mean UV energy density is
\begin{equation}
\begin{split}
{\cal E} ={a \over c}\left({3 Q(t) \over 4 \pi {\cal S}}
\right)^{1/3}n_0^{-4/3}
\int_0^{\infty} dM_{\rm dm} \int_0^t dt_{\rm f}\\
\times \Psi(M_{\rm dm},t_{\rm f}) \dot N(M_{\rm dm}, t_{\rm f}, t)
{\cal L}^{4/3}(M_{\rm dm}, t_{\rm f}, t).
\end{split}
\label{eq:ap4}
\end{equation}
The time evolution of the mean UV energy density is calculated solving
the energy conservation equation,
\begin{equation}
{d {\cal E} \over dt}= {\cal S} - \left[\gamma {\dot a \over a}
+x_{\HI}c \sigma_e n_H\right]{\cal E} - x_{\HII} {df_{II} \over dt}.
\label{eq:ap5}
\end{equation}
We have assumed that the spectral energy distribution of the ionising
radiation is $g_\nu \propto \nu^{-\gamma}$. Here $\sigma_e$ is the
effective UV energy loss cross section, defined as
$\sigma_e=(\Gamma_{\rm ph}+\Gamma_{\rm ion})/(4 \pi \overline J)$.  The time
derivative of the filling factor is $\dot f_{II}=(1-f_{II})\dot Q(t)$.
The porosity parameter $Q(t)$ and the volume filling factor
of the \HII regions are obtained solving \eq~(\ref{eq:ap5}) and
\eq~(\ref{eq:ap4}).

\subsection{Radiative Feedback}\label{ap:feed}

Feedback mechanisms on star formation are the most difficult part to
implement using the semianalytic approach. The recipes used should
be based on results of numerical simulations or scaling relations
based on observations (such as the dependence of the mass to light
ratio on the luminosity of galaxies). Analysing the simulations in
\cite{RicottiGSb:02}, it appears that star formation is regulated by the
interplay of several processes:
\begin{enumerate}
\item Internal feedback by UV radiation that produces photo-evaporative winds
\item Internal feedback by SN explosions
\item External (but local) feedback from UV radiation that regulates \H2
  formation/destruction and cooling
\item Thermal evolution of the IGM: the IGM Jeans mass sets the
  minimum mass of galaxies forming at a given redshift.
\end{enumerate}

In the semianalytic code, due to impossibility of modelling local
external feedback processes, we only implement process (iv) which is
global. We neglect the effects of internal feedbacks (i) and (ii). 

Galaxies with mass smaller than the Jeans mass in the IGM cannot
virialise because the pressure of the IGM prevents the gas from
falling into their halo potential well. Therefore the formation rate
of galaxies with masses $M_{\rm dm} < 10^8$ M$_\odot$ is suppressed if the
IGM is reionised or heated by X-ray or hard-UV background to $T
\approx 10,000$ K. This process works together with H$_2$ destruction
to suppress the formation of small-halo galaxies. If we take into
account the finite time required for pressure to influence the gas
distribution in the expanding universe, then the filtering mass,
$M_{\rm F, IGM}$, which depends on the full thermal history of the
IGM, provides a better fit to the simulation results than the Jeans
mass, which instead depends on instantaneous values of the sound speed
\citep{Gnedin:00b}.  The filtering mass of the IGM is simply related
to the Jeans mass by the relationship,
\begin{equation}
M_{\rm F, IGM}^{2/3}= {3 \over a}\int_0^a da^\prime M_J^{2/3}(a^\prime)\left[1-\left({a^\prime \over a}\right)^{1/2}\right].
\end{equation}
The Jeans mass is given by
\begin{equation}
M_J={4 \pi \over 3}\overline \rho \lambda_J^3,
\end{equation}
where $\lambda_J=2\pi c_s t_{\rm ff}$ is the Jeans length,
$t_{\rm ff} = (4 \pi G\overline \rho)^{-1/2}$ is the free-fall dynamical
time and $c_s$ is the IGM sound speed.

We implement the feedback imposing a minimum galaxy mass $M_{\rm min} =
M_{\rm F, IGM}(t_{\rm f})$ and computing the new SFR and the IGM temperature
iteratively. Using a geometric mean for the temperature we make sure
to achieve convergence. In practice we use a step-function kernel,
\begin{equation}
\Phi(M_{\rm dm}, t_{\rm f})= 
\begin{cases}
0, & \text{if $M_{\rm dm} < M_{\rm F, IGM}(t_{\rm f})$} \\
1, & \text{if $M_{\rm dm} \ge M_{\rm F, IGM}(t_{\rm f})$},
\end{cases}
\end{equation}
any time we integrate over the mass function of DM haloes.

\bibliographystyle{/home/ricotti/Latex/TeX/apj}
\bibliography{/home/ricotti/Latex/TeX/archive}

\label{lastpage}
\end{document}